\documentclass[a4paper,aip,jcp,preprint,superscriptaddress]{revtex4-2}
\usepackage{graphicx}
\usepackage{xspace,colortbl}
\usepackage{comment}
\usepackage{amsfonts}
\usepackage{amsmath}
\usepackage{amssymb}
\usepackage{rotating}
\usepackage{longtable,geometry}
\usepackage[english]{babel}
\usepackage[utf8]{inputenc}
\usepackage[babel]{csquotes}
\usepackage{listings}
\usepackage{centernot}
\usepackage{relsize}
\usepackage{stmaryrd}

\newcommand{\JCPformat}[4]{{#1} {\bf #2}, {#3} ({#4}).}

\renewcommand{\Ref}[4]{\JCPformat{#1}{#2}{#3}{#4}}

\newcommand{\tca}[3]{\Ref{Theor. Chim. Acta.}{#1}{#2}{#3}}

\newcommand{\jmathchem}[3]{\Ref{J. Math. Chem.}{#1}{#2}{#3}}

\newcommand{\molphys}[3]{\Ref{Mol. Phys.}{#1}{#2}{#3}}

\newcommand{\ijqc}[3]{\Ref{Int. J. Quantum Chem.}{#1}{#2}{#3}}

\newcommand{\physrev}[3]{\Ref{Phys. Rev.}{#1}{#2}{#3}}

\newcommand{\revmodphys}[3]{\Ref{Rev. Mod. Phys.}{#1}{#2}{#3}}

\renewcommand{\jcp}[3]{\Ref{J. Chem. Phys.}{#1}{#2}{#3}}

\def\a{\alpha}
\def\b{\beta}

\def\l{\lambda}

\def\s{\sigma}

\newcommand{\bt}{$\scriptstyle\blacktriangleright$ \ }
\newcommand{\overbar}[1]{\mkern 3mu\overline{\mkern-3mu#1\mkern-1mu}\mkern 1mu}

\DeclareFontFamily{U}{mathb}{\hyphenchar\font45}
\DeclareFontShape{U}{mathb}{m}{n}{
      <5> <6> <7> <8> <9> <10> gen * mathb
      <10.95> mathb10 <12> <14.4> <17.28> <20.74> <24.88> mathb12
      }{}
\DeclareSymbolFont{mathb}{U}{mathb}{m}{n}
\DeclareMathSymbol{\dlsh}{3}{mathb}{"EA}

\begin{document}

\title{2D-Block Geminals: a non 1-orthogonal and non 0-seniority model with reduced computational complexity}

\author{Patrick Cassam-Chena\"{\i}, Thomas Perez} 

\affiliation{Universit\'e C\^ote d'Azur, LJAD, UMR 7351, 06100 Nice, France}

\author{Davide Accomasso} 

\affiliation{Dipartimento di Chimica e Chimica Industriale, Universita` di Pisa, via Moruzzi 13, 56124 Pisa, Italy}

\begin{abstract}

We present a new geminal product wave function ansatz where the geminals are not constrained to be strongly orthogonal nor to be of seniority-zero. Instead, we introduce weaker orthogonality constraints between geminals which significantly lower the computational effort, without sacrificing the indistinguishability of the electrons. 
That is to say, the electron pairs corresponding to the geminals are not fully distinguishable, and their product has still to be antisymmetrized according to the Pauli principle to form a \textit{bona fide} electronic wave function.
Our geometrical constraints translate into simple equations involving the traces of products of our geminal matrices. In the simplest non-trivial model, a set of solutions is given by block-diagonal matrices where each block is of size 2x2 and consists of either a Pauli matrix or a normalized diagonal matrix, multiplied by a complex parameter to be optimized. With this simplified ansatz for geminals, the number of terms in the calculation of the matrix elements of quantum observables is considerably reduced. A proof of principle is reported and confirms that the ansatz is more accurate than strongly orthogonal geminal products while remaining computationally affordable.
\end{abstract}
\maketitle

Suggested running head:\\
2D-Block Geminals\\

All correspondance to be send to P. Cassam-Chena\"{\i},\\
cassam@unice.fr,\\

\newpage
\section{Introduction}

The accurate, \textit{ab initio} calculation of the electronic energy levels of a molecular system is still a challenging problem with exponential complexity with respect to the number of particles. This is particularly true for the so-called ``strongly correlated'' systems, where many Slater determinants have to be included in the wave function to achieve a qualitatively correct description of the molecular state. In such systems, it has been established that better ansätze can be based on an electron pair model, that is to say, an antisymmetric product of two-electron wave functions, called “geminals”.\\

However, without further restrictions, such a model has still a factorial computational cost with respect to the number of electronic orbitals and its applicability is therefore limited to small systems. Although an exponential scaling seems inherent in general fermionic quantum systems \cite{Troyer2005}, there is hope to achieve a better scaling for special classes of Fermionic systems. Reducing the scaling of geminal methods' computational cost  to a polynomial one is an active research field~\cite{Fecteau2022,Dutta2021,Wei2018,Tokmachev2016}. We postpone to the next section a review of the most popular proposals. Here, we just note that they often enforce the ``strong orthogonality'' constraint which breaks the indistinguishability of electron pairs, and is thus at odds with the Pauli principle, or in  non-orthogonal cases, they have been limited to seniority-zero wave functions with very few exceptions.\\ 

In the past, we have introduced generalized orthogonality constraints between geminals to reduce the computational effort, without sacrificing the indistinguishability of the electrons nor assuming a given seniority~\cite{Cassam2008-pr,Cassam2010}.
In other words, the electron pairs corresponding to the geminals were not fully distinguishable, so that  geminal products had still to be antisymmetrized according to the Pauli principle to form \textit{bona fide} electronic wave functions, and each geminal combined linearly  electronic configurations of mixed seniorities.  The aim of the present work is to pursue our exploration of orthogonality constraints to improve the computational affordability of  antisymmetric product of geminals (APG).\\

Recall that an APG wave function can be described in an alternative and  practical way, by a set of matrices, one for each geminal.  Our geometrical constraints translated in terms of these geminal matrices, into simple equations involving the traces of their products. Here, we will impose further geometrical constraints whose solutions will lead us to consider block-diagonal matrices, where each block is at most of size 2x2. These blocks are essentially Pauli or diagonal matrices, multiplied by a complex parameter to be optimized. With this simple ansatz for geminals, the number of terms in the calculation of  quantum observable matrix elements is considerably reduced.\\

The paper is organized as follows: In the next section we present an overview of existing APG-based methods. Then, we obtain an overlap formula for the general APG case, which exhibits its full complexity and serves us as a starting point to elaborate new constraints able to make it more amenable to practical computations. Next, we propose to enforce the  so-called “permutationally invariant $2$-orthogonality constraints” and show how we arrive at our 2D-block geminal model. Finally, we obtain the necessary quantities to implement our ansatz and provide a proof of principle that it is able to give strictly lower energies than strongly orthogonal geminal products. In the last section, we conclude on the prospect of our method.

%

\section{Antisymmetrized product of geminals (APG)  ansätze: an overview \label{review}}

Many APG models have appeared in the literature. We provide here a quick overview to help the reader in situating the model proposed in the following sections. We recall in Appendix A, the generalized concept of seniority,  and the definition of $p$-orthogonality, because these two notions make the constraints generally imposed on APG wave functions amenable to a synthetic presentation.

\subsection{General APG ansatz}

A general APG can be written as
\begin{equation}
 \Phi=g^1\wedge g^2\wedge\cdots\wedge g^p
 \label{AGP}
\end{equation}
with
\begin{equation}
 g^r =\sum_{i=1}^{2m-1}\sum_{j>i}^{2m} \lambda^r_{i,j}\, \varphi_{i}\wedge\varphi_{j} \quad.
 \label{gem}
\end{equation}
The $\varphi_i$'s form a basis of one-electron wavefunctions, not necessarily spin-adapted, i.e., they can be linear combinations involving both spin  $\alpha$ and spin $\beta$ spin-orbitals.
Each geminal $g^r$ is parametrized by $m(2m-1)$ independent scalar (real or complex) numbers $\lambda^r_{i,j}$.\\

However, even in the spin-adapted case, optimizing such ansatz is extremely costly from a computational standpoint \cite{McWeeny63} and has rarely been attempted \cite{Cassam2006,Cassam2007,Cassam2010}.
All popular geminal ansätze can be seen as APG with additional constraints. The latter can be divided into ``intra-geminal" constraints, bearing on the spin-orbitals of the geminal expansion Eq.(\ref{gem}), and ``inter-geminal" constraints, imposing  certain relationships between the geminals of the product Eq.(\ref{AGP}).\\

\subsection{Intra-geminal constraints}

\subsubsection{APsetG ansatz:}
A first approach consists in partitioning the spin-orbital basis into two subsets of cardinal $m$, $\mathcal{H}=\mathcal{H}_\a\oplus\mathcal{H}_\b$, ($dim\mathcal{H}_\a=dim\mathcal{H}_\b=m$), and to expand the geminals on two-electron basis functions of generalized seniority equal to $2$ with respect to this partition:
\begin{equation}
 g^r_{\text{set}} =\sum_{i=1}^{m}\sum_{j=1}^{m} \lambda^r_{i,j}\, \varphi_{2i-1}\wedge\varphi_{2j}, \quad \varphi_{2i-1}\in\mathcal{H}_\a, \quad \varphi_{2j}\in\mathcal{H}_\b,
\end{equation}
where the subscript ``set" hints to the fact that we have two disjoint sets of identical cardinal. One can choose for example to gather all spin-orbitals of spin $\alpha$ in the first set, $\mathcal{H}_\a$, and all those of spin $\beta$ in the second set, $\mathcal{H}_\b$. Then, one has wave functions that are eigenfunctions of $\hat{S}_z$ associated with the eigenvalue $0$.\\
Geminals of this type are parametrized by $m^2$ independent scalar numbers.\\

The Grassmann product of those geminals is called an APsetG (``Antisymmetric Product of set-divided Geminals'')~\cite{Johnson2017} wave function:
\begin{equation}
 \Phi_{\text{APsetG}} = \bigwedge_{r=1}^k g^r_{\text{set}} \quad.
\end{equation}\\

\subsubsection{APIG ansatz:}
An alternative idea is to partition the spin-orbital basis set into $m$ shells of the form $(\varphi_{2i-1},\varphi_{2i})$ and to build geminals out of seniority $0$ two-electron functions exclusively: \begin{equation}
 g^r_{\text{I}} =\sum_{i=1}^{m} \lambda^r_{i}\, \varphi_{2i-1}\wedge\varphi_{2i} \quad,
\end{equation}
where the subscript ``I'' stands for ``Interacting'', as the Grassmann product of those geminals is known as the APIG wave function (``Antisymmetric Product of Interacting Geminals'')~ \cite{Johnson2017}
\begin{equation}
 \Phi_{\text{APIG}} = \bigwedge_{r=1}^k g^r_{\text{I}} \quad.
\end{equation}
Each geminals in the product is parametrized by $m$ independent scalar numbers.\\

Note that, if the shells of the partition are such that, for all $i$, $\varphi_{2i-1}$ and $\varphi_{2i}$ are spin-orbitals of the same spatial part and of opposite spin, one retrieve the usual concept of seniority, and the APIG wave functions are those considered by Silver \cite{Silver69} and  revived recently by  Limacher~\cite{Limacher2016}. In this case a more natural notation is:
\begin{equation}
 g^r_{I} =\sum_{i=1}^{m} \lambda^r_{i}\, \varphi_{i}\wedge\overbar{\varphi_{i}} \quad.
\end{equation}\\

Most of other APG ansätze in use are particular cases of the APIG ansatz, hence of seniority $0$. Let us mention the antisymmetric product of rank-two geminals,
APr2G, where the coefficients are constrained to assume the form~\cite{Johnson2013,Tecmer2014}: $\lambda^r_{i}=\frac{1}{a_r\epsilon_i- b_i\lambda_r }$, $a_r,\epsilon_i,b_i,\lambda_r\in\mathbb{C}$. When these numbers are derived as eigenstate coefficients of reduced Bardeen-Cooper-Schrieffer (BCS) model Hamiltonians~\cite{Bardeen57}, one obtains the so-called  Richardson–Gaudin (RG) geminals which are currently under active development.~\cite{Johnson2020,Fecteau2022} 

\subsection{Inter-geminal constraints}

\subsubsection{AP1roG ansatz:}
Another important subcase of the  APIG model consists in imposing the following additional constraint: each geminal is only allowed to have a non-zero coefficient on a single pair of spin-orbitals (numbered from $1$ to $k$) pertaining to an  Hartree-Fock type reference wave function (with $k$ doubly occupied orbitals):
\begin{equation}
 g^r_{\text{1ro}} = \varphi_r\wedge\overbar{\varphi_r} + \sum_{i=k+1}^{m} \lambda^r_{i}\, \varphi_{i}\wedge\overbar{\varphi_{i}} \quad,
\end{equation}
where the subscript ``1ro" stands for ``1-reference orbital". Note that the constraint has an intra-geminal character but has also an inter-geminal one, since two different geminals must have different doubly occupied reference orbitals.
Each 1ro-geminal is parametrized by $m-k$ independent scalar numbers.\\

The Grassmann product of those geminals is called the ``AP1roG" wave function (``Antisymmetric Product of 1-reference orbital Geminals'')~\cite{Limacher2013,Limacher2014-mp,Boguslawski2014-prB,Boguslawski2014-jctc,Boguslawski2014-jcp,Tecmer2014}:
\begin{equation}
 \Phi_{\text{AP1roG}} = \bigwedge_{r=1}^k g^r_{\text{1ro}} \quad.
\end{equation}\\

\subsubsection{APSG ansatz:}
The $1$-orthogonality constraint (see Appendix \ref{ortho}) between the geminals in the Grassmann product has often been added as an inter-geminal condition. One obtains in this way an APSG (``Antisymmetrized Product of Strongly-orthogonal Geminals") wave functions, which have been extensively studied~\cite{Hurley53, Lykos56,Lykos56-err,Silver69,Mehler70,Silver70a,Silver70b,Silver70c,Rassolov02,Rassolov2004,Rassolov2007-1,Rassolov2007-jcp}. This ansatz, as well as its subcase, the generalized valence bond perfect-pairing (GVB-PP) wave function \cite{Goddard67,Goddard73}, have also been investigated for the calculation of excited states, with a time-dependent linear response theory approach~\cite{Chatterjee2015}.\\
The $1$-orthogonality  constraint reduces drastically the computational cost, however, it is usually too stringent to achieve chemical accuracy.\\

\subsubsection{AGP ansatz:}
At the other extreme, one can take all $k$ geminals, not just non-orthogonal but even equal. Then, one obtains a product wave function called ``AGP'' for ``Antisymmetrized Geminal Power'' \cite{Coleman97}. Dropping the superscript and denoting $g_{\text{AGP}}$ the geminal involved, we have:
\begin{equation}
\Phi_{\text{AGP}} = g_{\text{AGP}}^{\wedge k} = \underbrace{g_{\text{AGP}}\wedge\ldots\wedge g_{\text{AGP}}}_{\text{$k$ times}}\quad.
\end{equation}

\subsubsection{GMFCI p-orthogonal ans\"atze:}
The $p$-orthogonality constraint, $1\leq p\leq inf(n_1,n_2)$ between an $n_1$-electron wave function and an $n_2$-electron one (see Appendix \ref{ortho}), can be tuned from strong orthogonality ($p=1$), to ``weak orthogonality'' ($p=inf(n_1,n_2)$), to restrict variational freedom. Note that,
in the case ($n_1=n_2=n$), weak orthogonality ($p=n$) is just the usual orthogonality between $n$-electron wave functions.\\

Two such constraints, in between the APSG and AGP extreme cases, have been considered in the past for Geminal Mean Field Configuration Interaction (GMFCI) wave functions~\cite{Cassam2010}: (i) $2$-orthogonality between every pairs of geminals $(g^{r_1}, g^{r_2})$ of the APG; (ii) $2$-orthogonality between any geminal $g^{r}$ and any Grassmann product of the remaining ones $\bigwedge\limits_{s\neq r}g^{s}$.

\subsubsection{Various post-treatments:}

There have been many proposals to improve APSG or AP1roG wave functions by adding  perturbative  corrections~\cite{Rosta00,Rosta2002,Surjan2012,Jeszenszki2014,Cagg14-jcp,Limacher2014-pccp,Foldvari2019}. Other attempts have consisted in applying linearized coupled cluster corrections to APSG or AP1roG reference wave functions~\cite{Boguslawski2015,Zoboki2013}. However, these improvements were at the expense of the simplicity of the antisymmetrized geminal product model.\\ 

Many recent works aim at improving the AGP ansatz~\cite{Henderson2020,Khamoshi21}. A promising method called ``CJAGP'' for ``Cluster-Jastrow Antisymmetrized Geminal Power''~\cite{Neuscamman2013-CJAGP,Neuscamman2016-JCTC,Neuscamman2016-CJAGP}, or its ``JAGP'' particular case when atomic orbitals are not optimized~\cite{Neuscamman2012,Neuscamman2013-JAGP,Zhao2016-JCTC}, have been proposed by Neuscamman and can achieve polynomial cost, but with a large prefactor. 
Beside the latter approach qualified as ``substractive'' according to Neuscamman's interesting terminology, AGPs were also used as reference wave functions for more classical ``additive'' approaches~\cite{Kawasaki2016,Henderson2019}.\\

Note also that, very recently, the AGP ansatz~\cite{Khamoshi20} and an unitary version of AP1roG~\cite{Elfving21} have been ported to quantum computers.\\

In order to better understand the computational cost reductions occurring in the main ans\"atze just reviewed, and to propose new constrained  APG models, 
we find appropriate to return first to the general case in the next section. 

\section{General APG overlap formula}

In general, in quantum chemistry, the observables of interest, such as the Hamiltonian, contain at most two-particle operators. So, the computational cost of an electronic wave function method
is essentially given by the cost of calculating overlap integrals.\\

In this section, we provide a general formula, proved in Appendix \ref{proof},  for the overlap between two wave functions which are products of geminals, i.e. to compute $\langle \Phi_1\wedge\cdots\wedge\Phi_n|\Phi'_1\wedge\cdots\wedge\Phi'_n \rangle$, where the $\Phi_k$'s and the $\Phi'_k$'s are general geminals with $S_z=0$.\\

We will use a spin-adapted orthonormal basis set for the one-electron Hilbert space $\mathcal{H}$ (of finite dimension $2m$), that is made of two subsets $(\varphi_i)_{1\leq i\leq m}$ and $(\overbar{\varphi_i})_{1\leq i\leq m}$, the latter being, respectively,  spin $\alpha$ and spin $\beta$ spin-orbitals of the same set of orbitals.\\

We associate with each geminal, $\Phi_k$, a matrix of size $m\times m$, denoted $C_k$,  defined as follows:
\begin{equation}
\forall k\in\{1,\ldots,n\},\ C_k=\Big(\langle\varphi_{i}\wedge\overbar{\varphi_{j}}|\Phi_k\rangle\Big)_{\substack{1\leq i\leq m \\ 1\leq j\leq m}}\quad,
\end{equation}
so that:
\begin{equation}
 \forall k\in\{1,\ldots,n\},\ \Phi_k=\sum_{1\leq i,j\leq m} (C_k)_{i,j}\,\varphi_{i}\wedge\overbar{\varphi_{j}}\quad.
\end{equation}
Then the scalar product of two geminals can be expressed as the trace of the product of two matrices:
\begin{equation}
 \langle\Phi_i|\Phi_j\rangle=\text{tr}(C_i^\dagger C_j)\quad.
\end{equation}
Note that, the matrix associated with a singlet geminal, will be symmetric (even if we work with complex numbers), and that of a triplet geminal,  will be skew-symmetric.\\

Let $\Phi'_k$'s be geminals whose associated matrices are denoted by $C'_k$'s. As proved in Appendix \ref{proof}, the overlap between the wave functions $\Psi_e=\Phi_1\wedge\cdots\wedge\Phi_n$ and $\Psi'_e=\Phi'_1\wedge\cdots\wedge\Phi'_n$ is given by the formula:
\begin{equation} 
 \langle\Psi_e|\Psi'_e\rangle=\langle \Phi_1\wedge\cdots\wedge\Phi_n | \Phi'_1\wedge\cdots\wedge\Phi'_n \rangle=\sum_{\substack{0\leq N_{n,0},\ldots, N_{n,n}\leq n \\ \sum\limits_{k=0}^n N_{n,k}=\sum\limits_{k=0}^n k N_{n,k}=n}} (-1)^{N_{n,0}} \sum_{\sigma,\sigma'\in\mathfrak{S}_n} \prod_{k=1}^n \frac{\prod\limits_{j=1}^{N_{n,k}} T_{N_{n,0},\ldots,N_{n,k-1}}^{N_{n,k}} (j,\sigma,\sigma')}{k^{N_{n,k}}\, N_{n,k}!} 
 \label{ovgen1}
\end{equation}
with:
\begin{equation}
 \!\!\!\!\!\!\!\!\!\!\!\!\!\!\!T_{N_{n,0},\ldots,N_{n,k-1}}^{N_{n,k}} (j,\sigma,\sigma') = \text{tr}\Big[C_{\sigma\big(\sum\limits_{p=0}^{k-1} p N_{n,p}+(j-1)k+1\big)}^{\dagger}C'_{\sigma'\big(\sum\limits_{p=0}^{k-1} p N_{n,p}+(j-1)k+1\big)}\cdots \ C_{\sigma\big(\sum\limits_{p=0}^{k-1} p N_{n,p}+jk\big)}^{\dagger}C'_{\sigma'\big(\sum\limits_{p=0}^{k-1} p N_{n,p}+jk\big)}\Big]\ ,
  \label{ovgen2}
\end{equation}\\
if ${N_{n,k}}\ne 0$, and $T_{N_{n,0},\ldots,N_{n,k-1}}^{N_{n,k}} (j,\sigma,\sigma')=1$ by convention if ${N_{n,k}}= 0$. 

The external sum is over the "partitions of the integer $n$", i.e. the decompositions of $n$ into a sum of (possibly repeated) natural integers. Each such partition is entirely determined by a set of integers $(N_{n,k})_{k=1,\ldots, n}$, which correspond to the numbers of times integer $k$ is found in the decomposition. That is to say, we can write:
\begin{equation}
 n =\sum_{k=1}^n k N_{n,k}=\sum_{k=0}^n k N_{n,k} \quad.
\end{equation}
The number $N_{n,0}$, which is introduced in the second equality, extends the definition to $k=0$ and depends linearly on the others $N_{n,k}$'s:
\begin{equation}
 N_{n,0} = n - \sum_{k=1}^n N_{n,k} \quad.
\end{equation}
It plays a particularly important r\^ole in the formula, as it determines the sign factor of each term.\\

The internal sum on permutations  gives \textit{a priori} $(n!)^2$ terms. However, there are redundancies due to the invariance of a trace by circular permutation of its matrices, and by permutations of the traces of the same form, themselves. So, for a given partition of $n$, the number of \textit{a priori} distinct products of traces to be calculated, reduces to $\frac{(n!)^2}{\prod\limits_{k=1}^n k^{N_{n,k}}\, N_{n,k}!}$.\\

These products of traces are decomposed into double products, as follows:\\
The most external product runs over an index $k$  which determines the number of matrices  ($k$ matrices $C_i$ and $k$ matrices $C'_j$)  to be multiplied within each of the $N_{n,k}$ trace factors (provided $N_{n,k}\ne 0$) of the most internal product on the $j$-index.\\

\underline{\textbf{Example $(n=3)$ :}} We compute $\langle \Phi_1\wedge\Phi_2\wedge\Phi_3 |  \Phi'_1\wedge\Phi'_2\wedge\Phi'_3 \rangle=A_1+A_2+A_3$ , as follow:\\
\bt 3=0+0+3: $N_{3,0}=2,N_{3,1}=0,N_{3,2}=0,N_{3,3}=1$ $\leadsto$ $A_1=\sum\limits_{\sigma,\sigma'\in\mathfrak{S}_3}\frac{\text{tr}\Big[C_{\sigma(1)}^{\dagger}C'_{\sigma'(1)}C_{\sigma(2)}^{\dagger}C'_{\sigma'(2)}C_{\sigma(3)}^{\dagger}C'_{\sigma'(3)}\Big]}{3}$.\\
\bt 3=0+1+2: $N_{3,0}=1,N_{3,1}=1,N_{3,2}=1,N_{3,3}=0$ $\leadsto$ $A_2=-\sum\limits_{\sigma,\sigma'\in\mathfrak{S}_3}\frac{\text{tr}\Big[C_{\sigma(1)}^{\dagger}C'_{\sigma'(1)}\Big]\text{tr}\Big[C_{\sigma(2)}^{\dagger}C'_{\sigma'(2)}C_{\sigma(3)}^{\dagger}C'_{\sigma'(3)}\Big]}{2}$.\\
\bt 3=1+1+1: $N_{3,0}=0,N_{3,1}=3,N_{3,2}=0,N_{3,3}=0$ $\leadsto$ $A_3=\sum\limits_{\sigma,\sigma'\in\mathfrak{S}_3}\frac{\text{tr}\Big[C_{\sigma(1)}^{\dagger}C'_{\sigma'(1)}\Big]\text{tr}\Big[C_{\sigma(2)}^{\dagger}C'_{\sigma'(2)}\Big]\text{tr}\Big[C_{\sigma(3)}^{\dagger}C'_{\sigma'(3)}\Big]}{6}$.\\
 
Formula Eqs.(\ref{ovgen1}),(\ref{ovgen2}) can easily be checked in the particular case of an AGP against Eq.(8) of Ref.~\cite{Khamoshi19}. In this case, all geminal matrices are equal to one and the same matrix, say $C_0$:  $\forall k,\quad C_k=C'_k=C_0$. So, the sum over permutations gives just a factor $(n!)^2$. Moreover, by choosing the $\phi_i$'s to be the natural orbitals of the geminal function, we can make this matrix  diagonal, without loss of generality: $\forall i,j,\quad (C_0)_{i,j}=\delta_{i,j}\eta_i$. So, all traces reduce to traces of  powers of $C_0$, and $tr\Big[C_0^p\Big]=\sum\limits_{i=1}^m \eta_i^p$. In the above 
example, we obtain 
\begin{equation}\langle \Phi_1\wedge\Phi_2\wedge\Phi_3 |  \Phi_1\wedge\Phi_2\wedge\Phi_3 \rangle=36\times\Big(\frac{\sum\limits_{i=1}^m \eta_i^6}{3}-\frac{\sum\limits_{i=1}^m \eta_i^2\times\sum\limits_{j=1}^m \eta_j^4}{2}+\frac{\sum\limits_{i=1}^m \eta_i^2\times\sum\limits_{j=1}^m \eta_j^2\times\sum\limits_{k=1}^m \eta_k^2}{6}\Big).
\label{ovex}
\end{equation}
The coefficient of terms of the form $\eta_i^6$ is found to be $\frac{1}{3}-\frac{1}{2}+\frac{1}{6}=0$, and that of terms $\eta_i^2\eta_j^4$ with $i\ne j$ is $-\frac{1}{2}+3\times\frac{1}{6}=0$, so we are left with terms of the form $\eta_i^2\eta_j^2\eta_k^2$ with i,j,k all distinct, arising from the last term of Eq.(\ref{ovex}) only, with a combinatorial factor $3!$. Hence the final result:
\begin{equation}\langle \Phi_1\wedge\Phi_2\wedge\Phi_3 |  \Phi_1\wedge\Phi_2\wedge\Phi_3 \rangle=36\times\Big(\sum\limits_{1\leq i<j<k\leq m} \eta_i^2\eta_j^2\eta_k^2\Big) ,
\label{ovex2}
\end{equation}
which is the expected result, since in Ref.\cite{Khamoshi19}, both bra and ket AGP's are normalized by a factor $n!$, thus eliminating the factor $36=(3!)^2$. A similar path can be followed to retrieve the formula for an arbitrary $n$.\\

 At the other extreme, we can check Eqs.(\ref{ovgen1}),(\ref{ovgen2}) for the APSG case. The $1$-orthogonality constraint imposed to the geminals amounts to setting 
\begin{equation}
\forall i\ne j\quad C_i^{\dagger}C'_j=0.
\label{1-orth}
\end{equation}
So, only one partition of $n$ in the external sum gives a non-zero contribution, namely $n=\underbrace{1+\cdots +1}_{\text{$n$ times}}$. Moreover, this constraint also imposes $\sigma=\sigma'$ i.e. transforms the double sum on permutations into a single sum. The latter just gives a factor $N_{n,1}!=n!$ that cancels out with the redundancy factor in the denominator. Therefore, we obtain the expected formula:\\
\begin{equation} 
\langle \Phi_1\wedge\cdots\wedge\Phi_n | \Phi'_1\wedge\cdots\wedge\Phi'_n \rangle=\text{tr}\Big[C_{1}^{\dagger}C'_{1}\Big]\text{tr}\Big[C_{2}^{\dagger}C'_{2}\Big]\cdots\text{tr}\Big[C_{n}^{\dagger}C'_{n}\Big].
 \label{ovAPSG}
\end{equation}

One can also retrieve the more general formula recently obtained for APIG \cite{Moisset2022}, which encompasses the case of RG states first worked out by Sklyanin \cite{Sklyanin1999}. We were not aware of the latter work, when we first derived~\cite{Cassam2018-talk,Perez2019-poster} formula (\ref{ovgen1}), although it has a similar structure to the formula obtained by this author. In the cases of APIG, all matrices are diagonal, so their product is also a diagonal matrix whose i$^{th}$ diagonal element is just the product of the i$^{th}$ diagonal elements of the matrices in the product. Clearly, all matrices commute, so instead of intertwining bra and ket matrices in the traces, Eq.(\ref{ovgen2}), we can gather the bra matrices to the left, and the ket matrices to the right. So, our traces become exactly the sum over $i$ appearing in Eq.(69) of Ref.~\cite{Moisset2022}. In the latter equation there is also a sign factor and a numerical factor. After performing the product over the $r$ elements of a partition in Eq.(71) of Ref.~\cite{Moisset2022}  (where $M$ is $n$ in our notation), the sign factor reduces to $(-1)^{M-r}$, which is exactly the same as our global sign factor $(-1)^{N_{n,0}}$. Regarding the numerical factor, after performing the double summation on the symmetric group and the product over the $N_{n,k}$ $k$-cycles ($k$ in our notation is $q$ in  Ref.~\cite{Moisset2022}), we will obtain $N_{n,k}!\times(k!)^{2N_{n,k}}$ identical product of traces over $k$-cycles. The $N_{n,k}!$ comes from the commutations of the traces and simplifies with the denominator, and for each of the $N_{n,k}$ $k$-cycles
a factor $k!\times k!$ is due to the commutation of the $k$ bra matrices inside the trace, on one hand, and of the $k$-ket matrices on the other hand. Taking into account the $k^{N_{n,k}}$ in our denominator (related to the invariance by circular permutation of each trace), we obtain for each $k$-cycle a factor $k!\times (k-1)!$, which is exactly the factor in Eq.(69) of Ref.~\cite{Moisset2022}.

\section{Computational cost reduction  by orthogonality constraints}

The number of partitions of $n$ giving the number of terms in the external sum of Eq.(\ref{ovex}), grows asymptotically as
$\frac{1}{4n\sqrt{3}}exp\big(\pi\sqrt{\frac{2n}{3}}\big)$ when $n$ goes to infinity. So, even if one disregards the internal sum over  non-redundant permutations, the cost of calculating overlaps in the general case is rapidly prohibitive. Hence, the necessity of introducing constraints such as those reviewed in Section \ref{review}. Here, we investigate permutationally invariant $2$-orthogonality constraints already mentionned in conclusion of Ref.~\cite{ Cassam2010}.

\subsection{permutationally invariant 2-orthogonality constraints \label{PI2O}}
We have seen in the previous section how choosing all geminals parallel (actually equal) to one another (AGP) or all $1$-orthogonal to one another (APSG) simplifies the overlap formula.  Our aim is to propose  less drastic  inter-geminal orthogonality constraints, intermediate between the AGP  and the APSG  constraints.\\

Let us consider wave functions which are products of singlet or triplet geminals (i.e. with symmetric or antisymmetric associated matrices). We want to impose to the geminals $\Phi_k$'s what we call “permutationally invariant $2$-orthogonality constraints”:
\begin{equation}
 \forall i,j,k\in\{1,\ldots,n\} \ \text{distinct}, \ \left\{ \begin{array}{l l}
  \langle \Phi_i|\Phi_j\rangle=0\smallskip\\
  \Phi_k\dlsh(\Phi_i\wedge\Phi_j)=0
 \end{array} \right.
\end{equation}
(the interior product by $\Phi_k$, noted "$\Phi_k\dlsh$", plays in first quantization the same r\^ole as annihilation in second quantization) or, equivalently, in terms of matrices:
\begin{equation}
 \forall i,j,k\in\{1,\ldots,n\} \ \text{distinct}, \  \left\{ \begin{array}{l l}
  \text{tr}(C_i^\dagger C_j)=0\smallskip\\
  C_i C_k^\dagger C_j+C_j C_k^\dagger C_i=0
 \end{array} \right. \ .
 \label{perm-2-orth}
\end{equation}
Clearly Eq.(\ref{1-orth}) implies Eq.(\ref{perm-2-orth}) so $1$-orthogonality is at least as strong as permutationally invariant $2$-orthogonality. However, the question arises whether there actually exist geminals which are permutationally invariant $2$-orthogonal without being $1$-orthogonal i.e. whether $1$-orthogonality is strictly stronger than permutationally invariant $2$-orthogonality.\\

For $m=1$, there can only be one non-zero $C_i$ satisfying the $2$-orthogonal condition, so $1$-orthogonality is also satisfied. 
To answer the question positively, let us consider the case $m=2$, so that the $C_i$'s are $2\times 2$-matrices.   In this limit case, $\forall i, j, k\quad \text{distinct}, \quad \Phi_i\wedge\Phi_j\ne 0 \implies \Phi_i\wedge\Phi_j \propto \varphi_1\wedge\varphi_2\wedge\overbar{\varphi}_1\wedge\overbar{\varphi}_2 \implies \Phi_k=0$. So, to have at least three distinct matrices, we necessarily have $\forall i\ne j, \quad \Phi_i\wedge\Phi_j= 0$.  A general study~\cite{Perez2020} shows that one can have at most three singlet solutions of the form:\\
\begin{align}
 &C_1=\left( \begin{array}{cc}
   e^{-i\tau} & 0 \\
   0 & e^{i\tau} \\
  \end{array}  \right)z_1 \quad;\quad
 C_2=\left( \begin{array}{cc}
   \rho\,e^{-i\tau} & 1 \\
   1 & -\rho\,e^{i\tau} \\
  \end{array}  \right)z_2 \quad;\quad\\
 &C_3=\left( \begin{array}{cc}
   -e^{-i\tau} & \rho \\
   \rho & e^{i\tau} \\
  \end{array}  \right)z_3 \quad,\quad
  \text{with} \quad  \left\{ \begin{array}{c}
   \rho\in\mathbb{R}_+ \\
   \tau\in[0,\pi[
 \end{array}  \right.   \quad, \nonumber
\end{align}
and $z_1,z_2,z_3\in\mathbb{C^*}$ unimportant factors. To this set one can add at most one triplet geminal matrix. A convenient choice corresponding to $\rho=\tau=0, z_1=z_2=1, z_3=-1$ gives the set:
\begin{equation}
 I_2=\left( \begin{array}{cc}
   1 & 0 \\
   0 & 1 \\
  \end{array}  \right) \ ;\quad 
 \sigma_x=\left( \begin{array}{cc}
   0 & 1 \\
   1 & 0 \\
  \end{array}  \right) \ ;\quad 
 \sigma_z=\left( \begin{array}{cc}
   1 & 0 \\
   0 & -1 \\
  \end{array}  \right) \quad \text{and} \quad
 i\sigma_y=\left( \begin{array}{cc}
   0 & 1 \\
   -1 & 0 \\
  \end{array}  \right)\quad,
\end{equation}
where $\sigma_x$, $\sigma_y$ and $\sigma_z$ are the well-known Pauli matrices. We will call the set  $\{I_2, \sigma_x, i\sigma_y,\sigma_z\}$ the "$4$-type set". One can trivially verify that the associated geminals are not $1$-orthogonal.\\ 

Another non-maximal but useful set of solutions to the permutationally invariant $2$-orthogonality conditions in dimension 2 is $\{G_\theta,\sigma_x,i\sigma_y\}$, where:\\
\begin{equation}
 G_\theta = \left( \begin{array}{cc}
   \sqrt{2}\sin\theta & 0 \\
   0 & \sqrt{2}\cos\theta
  \end{array}  \right) \quad,\quad
 \text{with} \quad \theta\in[0,\pi[ \quad 
\end{equation}
whose associated geminal is:
\begin{equation}
 \Phi_{G_\theta} = \sqrt{2}\left(\sin\theta\,\varphi_1\wedge\overbar{\varphi_1} + \cos\theta\,\varphi_2\wedge\overbar{\varphi_2}\right) \quad \text{with} \quad ||\Phi_{G_\theta}|| =\sqrt{2} \quad.
\end{equation}
Note that by playing with $\theta$, one can retrieve special matrices of interest:
\begin{equation}
 \left\{ \begin{array}{c}
  G_{0} = \sqrt{2}\,E_{22}\smallskip\\
  G_{\frac{\pi}{4}} = I_2\smallskip\\
  G_{\frac{\pi}{2}} = \sqrt{2}\,E_{11}\smallskip\\
  G_{\frac{3\pi}{4}} = \sigma_z
 \end{array}\right. \quad,
\end{equation}
where $E_{ij}=(\delta_{ki}\delta_{lj})_{kl}$. It is important to also consider such "$3$-type sets" in order to break the degeneracy of the one-electron reduced density matrix (1RDM) eigenvalues. Note that all matrix types are either diagonal or anti-diagonal, so we partition them into two sets: $\mathcal{D}=\{I_2,\sigma_z,G_\theta\}$ and $\overbar{\mathcal{D}}=\{\sigma_x,i\sigma_y\}$.\\

One could study in the same fashion general solutions for increasing $m$-values. However, we prefer in the present article to limit ourselves to $m$-dimensional matrices built as block-diagonal matrices, each block being either a $1$-dimensional (1D) block ($1$-orthogonal part corresponding to a geminal $\varphi_i\wedge\overbar{\varphi_i}$) or a $2$-dimensional (2D) block constituted of a $2\times 2$-matrix  (whose elements correspond to geminal basis functions $\varphi_i\wedge\overbar{\varphi_i},\varphi_i\wedge\overbar{\varphi_{i+1}},\varphi_{i+1}\wedge\overbar{\varphi_i},\varphi_{i+1}\wedge\overbar{\varphi_{i+1}}$) belonging to the $4$-type set or to a $3$-type set for some value of $\theta$ to be optimized.\\

More explicitly, 
\begin{align} 
 &\quad\quad\quad\! C_i= \left( \begin{array}{ccccccc}
  A_i & & & & & &\\
  & \lambda_{\theta_1,i}^1\,B_{\theta_1,i}^1 & & & & &\\
  & & \ddots & & & &\\
  & & & \lambda_{\theta_{m''},i}^{m''}\,B_{\theta_{m''},i}^{m''} & & &\\
  & & & & \lambda_i^{m''+1}\,B_i^{m''+1} & &\\
  & & & & & \ddots &\\
  & & & & & & \lambda_i^{m'}\,B_i^{m'}
 \end{array}\right)\quad,\label{matmodfin} \\
 \nonumber  &\text{with} \quad 
 A_i = \left( \begin{array}{ccccccccc}
  0 & & & & & & & & \\
  & \ddots & & & & & & & \\
  & & 0 & & & & & & \\
  & & & \lambda_i^{-(h_{i-1}+1)} & & & & & \\
  & & & & \ddots & & & & \\
  & & & & & \lambda_i^{-h_i} & & & \\
  & & & & & & 0 & & \\
  & & & & & & & \ddots & \\
  & & & & & & & & 0\\
 \end{array}\right) \quad\text{and} \quad 
 \left\{ \begin{array}{c}
  \!\!\!B_{\theta_j,i}^j\in\{G_{\theta_j},\sigma_x,i\sigma_y\}\smallskip\\
  B_i^j\in\{I_2,\sigma_x,i\sigma_y,\sigma_z\}
 \end{array}\right. \quad,
\end{align}
the $\lambda_i^j$'s and the $\theta_k$'s being parameters to be optimized, $h_0=0$, and for $i>0, h_i-h_{i-1}$ is the dimension of the $1$-orthogonal subspace only populated in geminal $i$. The dimension of the total $1$-orthogonal subspace is thus $h_n$, and in this subspace, at most one geminal matrix can have a non-zero coefficient for a given $1D$-block, i.e. $\forall l\in\{1,\ldots, h_n\},\ i\ne j, \implies \lambda_i^{-l}\lambda_j^{-l}=0$. Note that we use negative integer superscripts for the $1D$-blocks, and positive superscripts for the $2D$-blocks. The number of $2\times2$ blocks is $m'=\frac{m-h_n}{2}\in\mathbb{N}$, the $m''$ first ones correspond to $3$-type-$2D$-blocks, the remaining $m'-m''$ ones to the $4$-type-$2D$-blocks.\\

Note that the permutationally invariant $2$-orthogonality conditions further impose that 
\begin{equation}
 \forall i,j,k\in\{1,\ldots,m\} \ \text{distinct}, \ \left\{ \begin{array}{l l}
  \sum\limits_{\overset{\text{$l$ such that }}{ B_i^l=B_j^l}}\overbar{\lambda_i^l}\lambda_j^l+ \sum\limits_{\overset{\text{$l$ such that }}{B_{\theta_l,i}^l=B_{\theta_l,j}^l}}\overbar{\lambda_{\theta_l,i}^l}\lambda_{\theta_l,j}^l=0\smallskip\\
  \lambda_i^l\overbar{\lambda_k^l}\lambda_j^l=0 \ \text{as soon as} \ \exists \{i',j'\}\subset\{i,j,k\}, \ B_{i'}^l=B_{j'}^l
 \end{array} \right. \quad.
\end{equation}
For practical reason, we choose to rather enforce the following sufficient set of conditions, called "extended permutationally invariant $2$-orthogonality" (EPI2O) conditions:
\begin{equation}
 \forall i,j\in\{1,\ldots,m\} \ \text{distinct}, \ \forall l\in\{1,\ldots,m'\}, \left\{\begin{array}{l l}  B_i^l=B_j^l \implies   \lambda_i^l\lambda_j^l=0 \\  B_{\theta_l,i}^l=B_{\theta_l,j}^l \implies   \lambda_{\theta_l,i}^l\lambda_{\theta_l,j}^l=0  \end{array} \right.\quad, \label{2econd}
\end{equation}
In other words, two distinct geminals are not allowed to have the same matrix type at a given block position, corresponding to superscript, say $l$. So, in a $4$-type $2D$-block,
there can be at most $4$ non-zero $\lambda_i^l$ coefficients, each in factor of one of the $4$ different block types, and similarly,  in a $3$-type $2D$-block, there can be at most $3$ non-zero $\lambda_{\theta_l,i}^l$ coefficient, each in factor of one of the $3$ possible block types.\\

Example:\\
Consider a linear $\text{H}_6$ system  with hydrogen nuclei equally spaced by one angstr\"om. An APG wave function of the $6$ electrons is a product of $3$ geminals:
\begin{equation}
 \Phi = \Phi_1\wedge\Phi_2\wedge\Phi_3.
\end{equation}
In a double zeta basis set,  if we do not include any $1D$-block to maximize deviation from $1$-orthogonality, the $12$ orbitals can be partitioned into six $2D$-blocks. Then, the $C_i$ matrices corresponding to the $\Phi_i$ geminals will have no $A_i$-sub-matrices.
Anticipating on the following, an optimized wave function within a model limited to spin-restricted $3$-type-$2D$-blocks,  is given by the following coefficients and sub-matrices in  APSG-optimized orbitals for a 6-31G basis set \cite{Hehre72}:\\
\begin{equation}
 C_1 = \left( \begin{array}{cccccc}
 0.699585\,G_{\theta_1} & & & & &\smallskip\\
 &  -0.021264\,G_{\theta_2} & & & & \smallskip\\
 & & 0.100629\,\sigma_x & & & \smallskip\\
 & & & 0 & & \smallskip\\
 & & & &  -0.001567\,\sigma_x & \smallskip\\
 & & & & & 0\smallskip\\
 \end{array}\right)
\end{equation}
\begin{equation}
 C_2 = \left( \begin{array}{cccccc}
0.113021\,\sigma_x & & & & &\smallskip\\
 &  0 & & & & \smallskip\\
 & &  0.697602\,G_{\theta_3} & & & \smallskip\\
 & & &  -0.024046\,G_{\theta_4} & & \smallskip\\
 & & & &  0 & \smallskip\\
 & & & & & -0.000019\,\sigma_x\smallskip\\
 \end{array}\right) 
\end{equation}
\begin{equation}
 C_3 = \left( \begin{array}{cccccc}
 0 & & & & &\smallskip\\
 &  -0.00076\,\sigma_x & & & & \smallskip\\
 & &  0 & & & \smallskip\\
 & & &  0.000016\,\sigma_x & & \smallskip\\
 & & & &  0.706656\,G_{\theta_5} & \smallskip\\
 & & & & & -0.025247\,G_{\theta_6}\smallskip\\
 \end{array}\right)
\end{equation}
the angles of the diagonal submatrices being:
\begin{equation}
 \left\{\begin{array}{l l}
  \theta_1 = 1.662992\ \text{rad}\smallskip\\
  \theta_2 = 1.391623\ \text{rad}\smallskip\\
  \theta_3 = 1.701670\ \text{rad}\smallskip\\
  \theta_4 = 1.267600\ \text{rad}\smallskip\\
  \theta_5 = 1.702433\ \text{rad}\smallskip\\
  \theta_6 = 1.268724\ \text{rad}\smallskip\\
 \end{array}\right. \quad.
\end{equation}

In the wave function representation, this translates into:
\begin{eqnarray}
\Phi_1&=&0.985161\times \varphi_1\wedge\overbar{\varphi_1}-0.091086\times\varphi_2\wedge\overbar{\varphi_2}-0.029590\times \varphi_3\wedge\overbar{\varphi_3}-0.005359\times \varphi_4\wedge\overbar{\varphi_4}\nonumber\\
&+&0.100629\times (\varphi_5\wedge\overbar{\varphi_6}+ \varphi_6\wedge\overbar{\varphi_5}) -0.001567\times (\varphi_{9}\wedge\overbar{\varphi_{10}}+\varphi_{10}\wedge\overbar{\varphi_{9}}),\nonumber
\end{eqnarray}
\begin{eqnarray}
\Phi_2&=&0.113021\times (\varphi_1\wedge\overbar{\varphi_2}+\varphi_2\wedge\overbar{\varphi_1})+0.9781214\times \varphi_5\wedge\overbar{\varphi_5}-0.128746\times \varphi_6\wedge\overbar{\varphi_6}\nonumber\\
&-&0.032455\times \varphi_7\wedge\overbar{\varphi_7}-0.010153\times \varphi_8\wedge\overbar{\varphi_8} -0.000019\times  (\varphi_{11}\wedge\overbar{\varphi_{12}}+\varphi_{12}\wedge\overbar{\varphi_{11}}),\nonumber
\end{eqnarray}
\begin{eqnarray}
\Phi_3&=&-0.00076\times (\varphi_3\wedge\overbar{\varphi_4}+\varphi_4\wedge\overbar{\varphi_3})+0.000016 \times (\varphi_7\wedge\overbar{\varphi_8}+ \varphi_8\wedge\overbar{\varphi_7})\nonumber\\
&-&0.990716\times \varphi_9\wedge\overbar{\varphi_9}-0.131173\times \varphi_{10}\wedge\overbar{\varphi_{10}} -0.034088\times  \varphi_{11}\wedge\overbar{\varphi_{11}}-0.010622\times\varphi_{12}\wedge\overbar{\varphi_{12}}.\nonumber
\end{eqnarray}
The expectation value of the energy for this APG wave function is found below the lowest APSG energy.\\

\textit{Remark 1}: A pair of successive $1D$-blocks associated with say $\varphi_i\wedge\overbar{\varphi_i},\varphi_{i+1}\wedge\overbar{\varphi_{i+1}}$ and occupied in a given geminal matrix, say $C_k$, can always be replaced by a $3$-type $2D$-block with  a non-zero coefficient $\lambda$ for this block in factor of a $G_\theta$ sub-matrix in $C_k$. So instead of optimizing two coefficients 
$\lambda_i$ and $\lambda_{i+1}$ in factor of the $1D$-blocks, one optimizes  $\lambda$ and  $\theta$, and  identifies  $\lambda_{i}=\lambda\times\sqrt{2}\sin\theta$, $\lambda_{i+1}=\lambda\times\sqrt{2}\cos\theta$. Conversely, in the previous example, if we neglect the term $+0.000016 \times (\varphi_7\wedge\overbar{\varphi_8}+ \varphi_8\wedge\overbar{\varphi_7})$ in $\Phi_3$, which is very small, then the fourth $2D$-block  could be replaced with two $1D$-blocks with non-zero coefficients in $\Phi_2$ only, taking the optimized values $-0.032455$ and $-0.010153$ respectively.

\textit{Remark 2}: In contrast, the coefficient of the seniority-$2$ term $+0.113021 \times (\varphi_1\wedge\overbar{\varphi_2}+ \varphi_2\wedge\overbar{\varphi_1})$ in $\Phi_2$, is quite large. This indicates that the introduction of a $\sigma_x$-submatrix in the first $2D$-block of $C_2$, which releases $1$-orthogonality between $\Phi_1$ and $\Phi_2$, improves significantly the APG wave function.

\subsection{Expression of the overlap for the EPI2O 2D-block geminal ansatz}

Let $\Phi_1,\ldots,\Phi_n$ and $\Phi'_1,\ldots,\Phi'_n$ be geminals verifying the EPI2O conditions with 
 for all $l$,  each $C'_k$ in the ket having the same matrix form as their $C_k$ counterpart in the bra (same block dimensions, types  and zero coefficients at the same places). One can anticipate a simplification of the overlap formula Eq.(\ref{ovgen1}) since in every block there will be at most $4$-cycle traces  (product of 4 pairs of block-sub-matrix) for the cases of spin-unrestricted $4$-type-$2D$-blocks, $3$-cycle traces for the cases of spin-restricted $4$-type-$2D$-blocks or spin-unrestricted $3$-type-$2D$-blocks, $2$-cycle traces for the cases of spin-restricted $3$-type-$2D$-blocks, or $1$-cycle traces for the cases of $1D$-blocks.\\

We shall prove that Eq.(\ref{ovgen1}) becomes:
\begin{equation}
 \langle \Phi_1\wedge\cdots\wedge\Phi_n | \Phi'_1\wedge\cdots\wedge\Phi'_n \rangle = \sum_{\substack{0\leq j_1,\ldots, j_n \leq m' \\ \text{distinct if non-zero} }} \zeta_{\Phi_1,\Phi'_1}^{j_1}\cdots\zeta_{\Phi_n,\Phi'_n}^{j_n}
 \label{ovEPI2O1}
\end{equation}
\begin{equation}
\text{with} \quad \zeta_{\Phi_u,\Phi'_u}^j=\left\{\begin{array}{c}
 \sum\limits_{t=h_{u-1}+1}^{h_u}\overbar{\lambda_u^{-t}}\lambda_u^{' -t} \quad \text{if}\ j=0 \medskip\\
 2\,\overbar{\lambda_{\theta_j,u}^j}\lambda_{\theta_j,u}^{' j} \quad \text{if}\ j\in\{1,\ldots, m''\}\medskip\\
 2\,\overbar{\lambda_u^j}\lambda_u^{' j} \quad \text{if}\  j\in\{m''+1,\ldots, m'\}\\
 \end{array}\right.
  \label{ovEPI2O2}
\end{equation}
\textbf{Direct demonstration:}\\
To prove the formula, we do not need to distinguish between the 
$3$-type-$2D$-blocks, and the $4$-type-$2D$-blocks, so we will omit the $\theta$ label of the coefficients and block matrices in the $3$-type cases. We decompose each geminal $\Phi_i$ into its $1$-orthogonal part and its EPI2O part:
\begin{equation}
 \Phi_i = \Phi_i^{0}+\sum_{x=1}^{m'} \Phi_i^x
\end{equation}
where
\begin{equation}
 \Phi_i^{0}:=\sum\limits_{t=h_{i-1}+1}^{h_i}\lambda_i^{-t}\,\varphi_t\wedge\overbar{\varphi_t} \quad,
\end{equation}
and
\begin{equation}
  \Phi_i^x=\sum_{p,q\in\{1,2\}}\lambda_i^x\big(B_i^x\big)_{p,q}\varphi_{\mathfrak{I}(x,p)}\wedge\overbar{\varphi_{\mathfrak{I}(x,q)}} \quad.
\end{equation}
$\mathfrak{I}$ denotes the function that maps component $1$ or $2$ of block $x\in\{1,\ldots,m'\}$ to the corresponding orbital index: 
\begin{equation}
    \left\{ \begin{array}{l }
 \mathfrak{I}(x,1)=h_n+2x-1\\
\mathfrak{I}(x,2)=h_n+2x\\
 \end{array} \right.\quad .
\end{equation}
Then,
\begin{equation}
 \!\!\!\!\!\!\!\!\!\!\!\!\!\!\!\!\!\!\!\!\!\!\!\!\bigwedge_{1\leq i\leq n}\Phi_i = \sum\limits_{k=1}^n\ \sum\limits_{1\leq i_1<\cdots<i_{n-k}\leq n}\ \sum_{1\leq x_1,\ldots,x_k\leq m'}\Phi_{i_1}^{0}\wedge\cdots\wedge\Phi_{i_{n-k}}^{0}\wedge\Phi_{\overbar{i_{1}}}^{x_1}\wedge\cdots\wedge\Phi_{\overbar{i_{k}}}^{x_k}\quad.
\end{equation}
where $\{i_1,\ldots,i_{n-k}\}\bigcup\{\overbar{i_{1}},\ldots,\overbar{i_{k}}\}= \{1,\ldots,n\}$, and $\overbar{i_{1}}<\ldots <\overbar{i_{k}}$. If the same $2D$-block is chosen, say $x_a=x_b$, for two different geminals $a$ and $b$, then the EPI2O conditions imply $B_{\overbar{i_{a}}}^{x_a}\ne B_{\overbar{i_{b}}}^{x_b}$, so that $\Phi_{\overbar{i_{a}}}^{x_a}\wedge\Phi_{\overbar{i_{b}}}^{x_b}=0$. Therefore, the internal sum can be restricted to $x_1,\ldots,x_k$ all distinct.\\

The $1$-orthogonality between distinct blocks, whether $1D$ or $2D$-blocks, implies that each term in the sum must be paired with its exact counterpart in $\bigwedge\limits_{1\leq i\leq n}\Phi'_i$ to give a non-zero contribution to the overlap. For the $1D$-part, since there is an exact correspondence between  geminal indices and non-zero $1D$-blocks, $1$-orthogonality is enough to associate a single partner for each block. In contrast, for a given $2D$-block, say $x$, there could be up to $4$ ket-geminal indices compatible with say $\Phi_{\overbar{i_{l}}}^{x}$. However, only the index corresponding to  the same index $\overbar{i_{l}}$ will give a non-zero contribution, since the EPI2O conditions imply  that different indices correspond to different matrix types and by $2$-orthogonality of the sub-matrices: $\overbar{i_{l}}\ne \overbar{i_{l'}} \implies \langle\Phi_{\overbar{i_{l}}}^{x}|\Phi_{\overbar{i_{l'}}}^{'\ x}\rangle=0$. So we are left with:
\begin{align*}
 \langle\bigwedge_{1\leq i\leq n}\Phi_i | \bigwedge_{1\leq i\leq n}\Phi'_i \rangle &=  \sum\limits_{k=1}^n\ \sum\limits_{1\leq i_1<\cdots<i_{n-k}\leq n}\ \sum_{\substack{1\leq x_1,\ldots,x_k\leq m'\\ \text{distinct}}}\prod\limits_{j=1}^{n-k} \langle\Phi_{i_j}^{0}|\Phi_{i_{j}}^{'\ 0}\rangle\prod\limits_{l=1}^{k}\langle\Phi_{\overbar{i_{l}}}^{x_l}|\Phi_{\overbar{i_{l}}}^{'\ x_l}\rangle\quad.
\end{align*}
Now, if we do not distinguish the different terms according to their number $k$ of $2D$-blocks, we obtain formulas (\ref{ovEPI2O1}) and (\ref{ovEPI2O2}) where the index $j$ of $\zeta_{\Phi_u,\Phi_u^{'j}}$ refers to a $1D$-overlap, $\langle \Phi_u^0| \Phi_u^{'0}\rangle $, if $j=0$, or to $2D$-overlap, $\langle \Phi_u^j| \Phi_u^{'j}\rangle$, if $j>0$.\\

In practice, it is often interesting to have $1D$-parts, to build a Hartree-Fock component in the geminal product wave function, for example, or to have less terms in the overlap formula. However, let us take advantage of Remark 1 to simplify the discussion and assume that we have only $2D$-blocks. Note that, if the number of orbital basis functions is odd, one can always add an extra one at infinity. Then  in Eq.(\ref{ovEPI2O1}), there will be only non-zero $j$-indices. Since the latter are necessarily all distinct,  Eq.(\ref{ovEPI2O1}) can be recast in the form:
\begin{eqnarray}
 \langle \Phi_1\wedge\cdots\wedge\Phi_n | \Phi'_1\wedge\cdots\wedge\Phi'_n \rangle &=& \sum_{1\leq j_1<\ldots < j_n \leq m'} \sum\limits_{\sigma\in\mathfrak{S}_n} \zeta_{\Phi_1,\Phi'_1}^{\sigma(j_1)}\cdots\zeta_{\Phi_n,\Phi'_n}^{\sigma(j_n)}\nonumber\\
 &=& \sum_{1\leq j_1<\ldots < j_n \leq m'} \sum\limits_{\sigma\in\mathfrak{S}_n} \zeta_{\Phi_{\sigma(1)},\Phi'_{\sigma(1)}}^{j_1}\cdots\zeta_{\Phi_{\sigma(n)},\Phi'_{\sigma(n)}}^{j_n}\nonumber\\
  &=&  \sum\limits_{\sigma\in\mathfrak{S}_n} \sum_{1\leq j_1<\ldots < j_n \leq m'} \zeta_{\Phi_{\sigma(1)},\Phi'_{\sigma(1)}}^{j_1}\cdots\zeta_{\Phi_{\sigma(n)},\Phi'_{\sigma(n)}}^{j_n},
 \label{ovEPI2O3}
\end{eqnarray}
where $\sigma$ runs over all permutations of $n$-indices, $\mathfrak{S}_n$. For a block $x$, let $\mathcal{U}(x)=\{u\in\{1,\ldots,n\},\ \lambda_u^x\ne 0\}$ be the set of indices of the geminals which contain a submatrix with a non-zero coefficient for this block.
In fact, in Eq.(\ref{ovEPI2O3}) there are always a large number of terms that are zero, since  the permutation  must fulfill $\forall i\in\{1,\ldots,n\},\ \sigma(i)\in\mathcal{U}(j_i)$ otherwise $\zeta_{\Phi_{\sigma(i)},\Phi'_{\sigma(i)}}^{j_i}=0$ and we recall that $Card(\mathcal{U}(j_i))\le 4$.\\

Let us set $\mathcal{I}_{n}^{m'}(\sigma)=\sum\limits_{1\leq x_1<\ldots < x_n \leq m'} \zeta_{\Phi_{\sigma(1)},\Phi'_{\sigma(1)}}^{x_1}\cdots\zeta_{\Phi_{\sigma(n)},\Phi'_{\sigma(n)}}^{x_n}$ and more generally
for $p$ indices $u_1<\cdots<u_p$ , and $q$ block indices $y_1<\cdots<y_q$, let us denote $\mathcal{I}_{n\smallsetminus\{u_1,\ldots,u_p\}}^{m'\smallsetminus\{y_1,\ldots,y_q\}}(\sigma)$ the quantity:
\begin{equation*}
 \!\!\!\!\!\!\sum\limits_{\substack{1\leq x_1<\cdots<x_{n-p}\leq m' \\  x_1,\ldots,x_{n-p}\notin\{y_1,\ldots,y_q\}}}\!\!\!\!\!\!\!\!\zeta_{\Phi_{\sigma(1)},\Phi'_{\sigma(1)}}^{x_1}\cdots\zeta_{\Phi_{\sigma(u_1-1)},\Phi'_{\sigma(u_1-1)}}^{x_{u_1-1}}\zeta_{\Phi_{\sigma(u_1+1)},\Phi'_{\sigma(u_1+1)}}^{x_{u_1}}\!\!\cdots\zeta_{\Phi_{\sigma(u_p-1)},\Phi'_{\sigma(u_p-1)}}^{x_{u_p-p}}\zeta_{\Phi_{\sigma(u_p+1)},\Phi'_{\sigma(u_p+1)}}^{x_{u_p+1-p}}\!\!\cdots\zeta_{\Phi_{\sigma(n)},\Phi'_{\sigma(n)}}^{x_{n-p}}
\end{equation*}
where $q$ geminals and $p$ blocks are excluded with respect to $\mathcal{I}_{n}^{m'}(\sigma)$. 
This allows us to write the following recursion formula: 
\begin{equation}
 \mathcal{I}_{n}^{m'}(\sigma)=\mathcal{I}_{n}^{m'\smallsetminus\{y\}}(\sigma)+\sum\limits_{\substack{k=1\\ \sigma(k)\in\mathcal{U}(y)}}^n\zeta_{\Phi_{\sigma(k)},\Phi'_{\sigma(k)}}^{y}\times \mathcal{I}_{n\smallsetminus\{k\}}^{m'\smallsetminus\{y\}}(\sigma),
\end{equation}
which can be useful to compute the $\mathcal{I}_{n}^{m'}(\sigma)$'s efficiently.\\

\subsection{Expression of the 2RDM for the EPI2O 2D-block geminal ansatz}

To optimize variationally the expectation value of the electronic Coulomb Hamiltonian, we will make use of its contracted form
\begin{equation}
^2\hat{h}=\frac{1}{n-1}\left(\hat{T}_1 + \hat{T}_2 + \sum_i Z_i\times\left( \frac{1}{\Vert \hat{R}_i-\hat{r}_1\Vert}+\frac{1}{\Vert \hat{R}_i-\hat{r}_2\Vert}\right) \right) +\frac{1}{\Vert \hat{r}_1-\hat{r}_2\Vert},
\end{equation} 
where $\hat{T}_1$, $\hat{T}_2$ and $\hat{r}_1$, $\hat{r}_2$ are, respectively, the kinetic and position operators of electrons $1$ and $2$,  $Z_i$ the charge of nucleus $i$, and $\hat{R}_i$ the position operator of nucleus $i$. The expectation value over wave function $\Psi_e$ is then computed according to:
\begin{equation}
E=\text{tr}(^2\Gamma(\Psi_e)\  ^2\hat{h}).
\end{equation}  
So,  only  the $2$-electron reduced density matrix (2RDM), $^2\Gamma(\Psi_e)$ is needed \textit{in fine}.\\

In order to provide general expressions for the more general, $2$-electron, reduced, transition matrix ($2RTM$) between wave functions $\Psi_e$ and $\Psi'_e$, that we will denote  $^2\Gamma(\Psi_e,\Psi'_e)$ (the $2RDM$ expression will be obtained by suppressing the primes), let us generalize our notation to $1D$-blocks (using negative integers to number them):
\begin{equation*}
\forall i\in\{1,\ldots,n\},\quad \forall t\in\{h_{i-1}+1,\ldots,h_i\},\ \text{we set}\  \Phi_i^{-t}=\lambda_i^{-t}\,\varphi_t\wedge\overbar{\varphi_t}\ \text{and}\  
\mathcal{U}(-t)=\{i\}.
\end{equation*}
For any block $j$, we note $\overbar{\mathcal{U}}(j)$ the complement of   $\mathcal{U}(j)$ in $\{1,\ldots,n\}$.
We also set for $p$ geminal indices, $1\leq u_1<\cdots<u_p\leq n$, and $q$ $2D$-block indices\\ $0<y_1<\cdots<y_q<m'$, 
\begin{eqnarray}
\lefteqn{\mathcal{S}_{n\smallsetminus\{u_1,\ldots,u_p\}}^{m'\smallsetminus\{y_1,\ldots,y_q\}}=}\nonumber\\
 &\!\!\!\!\!\!\sum\limits_{\substack{0\leq x_1,\cdots,x_{n-p}\leq m' \\\text{all distinct if non-zero}\\  x_1,\ldots,x_{n-p}\notin\{y_1,\ldots,y_q\}}}\!\!\!\!\!\!\!\!\zeta_{\Phi_{1},\Phi'_{1}}^{x_1}\cdots\zeta_{\Phi_{u_1-1},\Phi'_{u_1-1}}^{x_{u_1-1}}\zeta_{\Phi_{u_1+1},\Phi'_{u_1+1}}^{x_{u_1}}\!\!\cdots\zeta_{\Phi_{u_p-1},\Phi'_{u_p-1}}^{x_{u_p-p}}\zeta_{\Phi_{u_p+1},\Phi'_{u_p+1}}^{x_{u_p+1-p}}\!\!\cdots\zeta_{\Phi_{n},\Phi'_{n}}^{x_{n-p}} \ . &
\end{eqnarray}
 Since $\Psi_e$ and $\Psi'_e$ are product of $(S_z=0)$-geminals, the $2RTM$ can be decomposed into $4$ $\hat{S}_z$-adapted blocks that we denote simply (omitting the $(\Psi_e,\Psi'_e)$ dependency) $^2\Gamma^{\gamma,\delta}$ with $\gamma,\delta\in\{\alpha,\beta\}$ referring to spin operator $\hat{S}_z$ eigenstates.\\
 
 Let us first consider $^2\Gamma^{\alpha,\beta}$, the $^2\Gamma^{\beta,\alpha}$ block being obtained by symmetry.
 Denoting $\mathbf{c}_i$ (respectively  $\mathbf{\bar{c}}_{i}$)  the annihilation operator of spin-orbital $\varphi_i$ (respectively $\overbar{\varphi}_{i}$)
\begin{align*}
&\forall i_1,j_1,i_2,j_2\in\{1,\ldots,m\},\\
^2\Gamma^{\alpha,\beta}_{i_1j_1,i_2j_2}&= \langle \mathbf{\bar{c}}_{j_1}\mathbf{c}_{i_1}\Psi_e|  \mathbf{\bar{c}}_{j_2}\mathbf{c}_{i_2}\Psi'_e \rangle\quad \Leftrightarrow&\\
^2\Gamma^{\alpha,\beta}_{i_1j_1,i_2j_2}&= \langle \varphi_{i_1}\wedge\overbar{\varphi}_{j_1}\dlsh \Phi_1\wedge\cdots\wedge\Phi_n| \varphi_{i_2}\wedge \overbar{\varphi}_{j_2}\dlsh\Phi'_1\wedge\cdots\wedge\Phi'_n \rangle&\\
&= \sum\limits_{\substack{-h_n\leq x_1,\ldots,x_n\leq m'\\
\text{all distinct}\\ 
-h_n\leq y_1,\ldots,y_n\leq m'\\
\text{all distinct}}}
\!\!\!\!\!\!\langle \varphi_{i_1}\wedge\overbar{\varphi}_{j_1}\dlsh \Phi_{1}^{x_1}\wedge\cdots\wedge\Phi_{n}^{x_n}| \varphi_{i_2}\wedge \overbar{\varphi}_{j_2}\dlsh\Phi_{1}^{'\ y_1}\wedge\cdots\wedge\Phi_{n}^{'\ y_n} \rangle.&
\end{align*}
Note that the summation can be restricted to block $x_i$ and $y_i$ such that $i\in\mathcal{U}(x_i)\cap\mathcal{U}(y_i)$ for all $i\in\{1,\ldots,n\}$.
We define the "block function", $\mathfrak{B}$, that associates to an orbital its block number: 
\begin{align*}
 \mathfrak{B} : \quad \{1,\ldots,m\} &\longrightarrow \mathbb{Z}^*\\
 i &\longmapsto \mathfrak{B}(i)=\left\{ \begin{array}{l l}
 -i & \ \text{if} \ i\leq h_n\\
 \left\lfloor \frac{i-h_n+1}{2} \right\rfloor & \ \text{if} \ i>h_n
 \end{array} \right.\quad.
\end{align*}
where $\left\lfloor z \right\rfloor$ is the floor of $z$, and the function $\mathfrak{C}$, that associates to an orbital of a $2D$-block its position in the block: 
\begin{align*}
 \mathfrak{C} : \quad \{h_n+1,\ldots,m\} &\longrightarrow \{1,2\}\\
 i &\longmapsto \mathfrak{C}(i)= i-h_n-2(\mathfrak{B}(i)-1).
\end{align*}
 Finally, we introduce the notation $\bar{j}$ for the complementary of index $j\in\{1,2\}$, that is to say,  $\bar{j}=2$ if $j=1$, and $\bar{j}=1$ if $j=2$. Then for orbital $i$ and geminal $k$, we associate the index $\mathfrak{D}(i,k)$ defined as follow:
\begin{align*}
\left\{ \begin{array}{l l}
 \mathfrak{D}(i,k)= \mathfrak{C}(i)& \ \text{if} \ B_{k}^{\mathfrak{B}(i)}\in\mathcal{D}(\mathfrak{B}(i))\\
  \mathfrak{D}(i,k)= \overbar{\mathfrak{C}(i)}& \ \text{if} \ B_{k}^{\mathfrak{B}(i)}\in\overbar{\mathcal{D}}(\mathfrak{B}(i))\\
 \end{array} \right.\quad.
\end{align*}\\

\underline{\textbf{Case 1: $\mathfrak{B}(i_1)\ne\mathfrak{B}(j_1)$}}\\
Two distinct blocks are broken in the bra, then, to have at least one ket with non-zero overlap, it is necessary that $\mathfrak{B}(i_2)=\mathfrak{B}(i_1)$ and $\mathfrak{B}(j_2)=\mathfrak{B}(j_1)$. Moreover, these two blocks have to appear within both $x_1,\ldots,x_n$ and  $y_1,\ldots,y_n$. The remaining $n-2$ blocks in each set must be two by two identical because of $1$-orthogonality between the blocks, and each pair must be associated with the same geminal because of the EPI2O conditions. This means also that the $2$ broken blocks  must be associated with the same pair of geminals in the bra and in the ket,  of indices say $(k_1,k_2)$.\\

\underline{Subcase 1a: $\mathfrak{B}(i_1)<0$ and $\mathfrak{B}(j_1)<0$}\\
Then, necessarily $i_2=i_1$, $j_2=j_1$ and $\mathcal{U}(\mathfrak{B}(i_1))\ne\mathcal{U}(\mathfrak{B}(j_1))$  otherwise $^2\Gamma^{\alpha,\beta}_{i_1j_1,i_2j_2}=0$. $k_1$ is the unique element of $\mathcal{U}(\mathfrak{B}(i_1))$ and  $k_2$ is the unique element of $\mathcal{U}(\mathfrak{B}(j_1))$ 
\begin{equation}
^2\Gamma^{\alpha,\beta}_{i_1j_1,i_1j_1}=\overbar{\lambda}_{k_1}^{-i_1}\lambda_{k_1}^{'\ -i_1}\overbar{\lambda}_{k_2}^{-j_1}\lambda_{k_2}^{'\ -j_1}\mathcal{S}_{n\smallsetminus\{k_1,k_2\}}^{m'}.
\label{1a-ab}
\end{equation}

\underline{Subcase 1b: $\mathfrak{B}(i_1)<0$ and $\mathfrak{B}(j_1)>0$}\\
Then $^2\Gamma^{\alpha,\beta}_{i_1j_1,i_2j_2}\ne 0$, implies $i_2=i_1$ and $k_1$ is the unique element of $\mathcal{U}(\mathfrak{B}(i_1))$. The block $\mathfrak{B}(j_1)$ is associated with the same geminal $k_2\in\mathcal{U}(\mathfrak{B}(j_1))$ in the bra and in the ket.
The block matrix $B_{k_2}^{\mathfrak{B}(j_1)}$ being either diagonal or anti-diagonal, after annihilation of the $\beta$-spin-orbitals, one can only match the lone ket and bra $\alpha$-spin-orbitals if $j_2=j_1$. Then:
\begin{eqnarray}
^2\Gamma^{\alpha,\beta}_{i_1j_1,i_1j_1}&=&\sum\limits_{k_2\in\mathcal{U}(\mathfrak{B}(j_1))}\overbar{\lambda}_{k_1}^{-i_1}\lambda_{k_1}^{'\ -i_1}\overbar{\lambda}_{k_2}^{\mathfrak{B}(j_1)}\lambda_{k_2}^{'\ \mathfrak{B}(j_1)}(B_{k_2}^{\mathfrak{B}(j_1)})_{\mathfrak{D}(j_1,k2),\mathfrak{C}(j_1)}^2\mathcal{S}_{n\smallsetminus\{k_1,k_2\}}^{m'\smallsetminus\{\mathfrak{B}(j_1)\}}\nonumber
\end{eqnarray}
which can be decomposed as:
\begin{eqnarray}
^2\Gamma^{\alpha,\beta}_{i_1j_1,i_1j_1}&=&\sum\limits_{k_2\in\mathcal{D}(\mathfrak{B}(j_1))}\overbar{\lambda}_{k_1}^{-i_1}\lambda_{k_1}^{'\ -i_1}\overbar{\lambda}_{k_2}^{\mathfrak{B}(j_1)}\lambda_{k_2}^{'\ \mathfrak{B}(j_1)}(B_{k_2}^{\mathfrak{B}(j_1)})_{\mathfrak{C}(j_1),\mathfrak{C}(j_1)}^2\mathcal{S}_{n\smallsetminus\{k_1,k_2\}}^{m'\smallsetminus\{\mathfrak{B}(j_1)\}}\nonumber\\
&+&\sum\limits_{k_2\in\overbar{\mathcal{D}}(\mathfrak{B}(j_1))}\overbar{\lambda}_{k_1}^{-i_1}\lambda_{k_1}^{'\ -i_1}\overbar{\lambda}_{k_2}^{\mathfrak{B}(j_1)}\lambda_{k_2}^{'\ \mathfrak{B}(j_1)}\mathcal{S}_{n\smallsetminus\{k_1,k_2\}}^{m'\smallsetminus\{\mathfrak{B}(j_1)\}}.
\label{1b-ab}
\end{eqnarray}
where $\mathcal{D}(j)$ (resp. $\overbar{\mathcal{D}}(j)$) is the subset of $\mathcal{U}(j)$ collecting only the indices corresponding to geminals having a diagonal (resp. anti-diagonal) submatrix for block $j$.  $\forall j\in\{1,\ldots,m'\}, \  \mathcal{U}(j)=\mathcal{D}(j)\bigcup \overbar{\mathcal{D}}(j)$. Note that the factor $(B_{k_2}^{\mathfrak{B}(j_1)})_{\mathfrak{C}(j_1),\mathfrak{C}(j_1)}^2$ on the first line of Eq.(\ref{1b-ab}) will be $1$ in the $4$-type case, and either $2cos^2\theta$ or $2sin^2\theta$ according to $\mathfrak{C}(j_1)$ being $1$ or $2$ in the $3$-type case. On the second line, we used the fact that $\forall k_2\in\overbar{\mathcal{D}}(\mathfrak{B}(j_1)),\quad(B_{k_2}^{\mathfrak{B}(j_1)})_{1,2}^2=(B_{k_2}^{\mathfrak{B}(j_1)})_{2,1}^2=1$. 

\underline{Subcase 1c: $\mathfrak{B}(i_1)>0$ and $\mathfrak{B}(j_1)<0$}\\
By symmetry with the previous subcase:
\begin{eqnarray}
^2\Gamma^{\alpha,\beta}_{i_1j_1,i_1j_1}&=&\sum\limits_{k_1\in\mathcal{D}(\mathfrak{B}(i_1))}\overbar{\lambda}_{k_1}^{\mathfrak{B}(i_1)}\lambda_{k_1}^{'\ \mathfrak{B}(i_1)}(B_{k_1}^{\mathfrak{B}(i_1)})_{\mathfrak{C}(i_1),\mathfrak{C}(i_1)}^2\overbar{\lambda}_{k_2}^{-j_1}\lambda_{k_2}^{'\ -j_1}\mathcal{S}_{n\smallsetminus\{k_1,k_2\}}^{m'\smallsetminus\{\mathfrak{B}(i_1)\}}\nonumber\\
&+&\sum\limits_{k_1\in\overbar{\mathcal{D}}(\mathfrak{B}(i_1))}\overbar{\lambda}_{k_1}^{\mathfrak{B}(i_1)}\lambda_{k_1}^{'\ \mathfrak{B}(i_1)}\overbar{\lambda}_{k_2}^{-j_1}\lambda_{k_2}^{'\ -j_1}\mathcal{S}_{n\smallsetminus\{k_1,k_2\}}^{m'\smallsetminus\{\mathfrak{B}(i_1)\}}.
\label{1c-ab}
\end{eqnarray}

\underline{Subcase 1d: $\mathfrak{B}(i_1)>0$ and $\mathfrak{B}(j_1)>0$}\\
This last case is more complicated than the previous ones, since there are $3$ sub-subcases to consider:\\
(i) $(k_1,k_2)\in\mathcal{U}(\mathfrak{B}(i_1))\bigcap\mathcal{U}(\mathfrak{B}(j_1))\times\mathcal{U}(\mathfrak{B}(i_1))\bigcap\mathcal{U}(\mathfrak{B}(j_1))$,\\
(ii) $(k_1,k_2)\in\mathcal{U}(\mathfrak{B}(i_1))\times\overbar{\mathcal{U}}(\mathfrak{B}(i_1))\bigcap\mathcal{U}(\mathfrak{B}(j_1))$,\\
(iii) $(k_1,k_2)\in\mathcal{U}(\mathfrak{B}(j_1))\times\mathcal{U}(\mathfrak{B}(i_1))\bigcap\overbar{\mathcal{U}}(\mathfrak{B}(j_1))$.\\
Using the notation already defined, the following general formula can be obtained for case (i):

\begin{eqnarray}
^2\Gamma^{\alpha,\beta}_{i_1j_1,i_2j_2}&=&\!\!\!\!\!\!\sum\limits_{\substack{k_1,k_2\in\mathcal{U}(\mathfrak{B}(i_1))\bigcap\mathcal{U}(\mathfrak{B}(j_1))\\ k_1\ne k_2}}\!\!\!\!\!\!\mathcal{S}_{n\smallsetminus\{k_1,k_2\}}^{m'\smallsetminus\{\mathfrak{B}(i_1),\mathfrak{B}(j_1)\}}\left(\delta_{i_1,i_2}\delta_{j_1,j_2}\left(\overbar{\lambda}_{k_1}^{\mathfrak{B}(i_1)}\lambda_{k_1}^{'\ \mathfrak{B}(i_1)}(B_{k_1}^{\mathfrak{B}(i_1)})_{\mathfrak{C}(i_1),\mathfrak{D}(i_1,k_1)}^2\overbar{\lambda}_{k_2}^{\mathfrak{B}(j_1)}\lambda_{k_2}^{'\ \mathfrak{B}(j_1)}\right.\right.\nonumber\\
&\times&\left.(B_{k_2}^{\mathfrak{B}(j_1)})_{\mathfrak{D}(j_1,k_2),\mathfrak{C}(j_1)}^2+\overbar{\lambda}_{k_2}^{\mathfrak{B}(i_1)}\lambda_{k_2}^{'\ \mathfrak{B}(i_1)}(B_{k_2}^{\mathfrak{B}(i_1)})_{\mathfrak{C}(i_1),\mathfrak{D}(i_1,k_2)}^2\overbar{\lambda}_{k_1}^{\mathfrak{B}(j_1)}\lambda_{k_1}^{'\ \mathfrak{B}(j_1)}(B_{k_1}^{\mathfrak{B}(j_1)})_{\mathfrak{D}(j_1,k_1),\mathfrak{C}(j_1)}^2\right)\nonumber\\
&+&\delta_{\mathfrak{D}(i_1,k_1),\mathfrak{D}(i_2,k_2)}\delta_{\mathfrak{D}(j_1,k_2),\mathfrak{D}(j_2,k_1)}\left(\overbar{\lambda}_{k_1}^{\mathfrak{B}(i_1)}\lambda_{k_2}^{'\ \mathfrak{B}(i_1)}(B_{k_1}^{\mathfrak{B}(i_1)})_{\mathfrak{C}(i_1),\mathfrak{D}(i_1,k_1)}(B_{k_2}^{\mathfrak{B}(i_1)})_{\mathfrak{C}(i_2),\mathfrak{D}(i_2,k_2)}\right.\nonumber\\
&\times&\left.\overbar{\lambda}_{k_2}^{\mathfrak{B}(j_1)}\lambda_{k_1}^{'\ \mathfrak{B}(j_1)}(B_{k_2}^{\mathfrak{B}(j_1)})_{\mathfrak{D}(j_1,k_2),\mathfrak{C}(j_1)}B_{k_1}^{\mathfrak{B}(j_1)})_{\mathfrak{D}(j_2,k_1),\mathfrak{C}(j_2)}\right)\nonumber\\
&+& \delta_{\mathfrak{D}(i_1,k_2),\mathfrak{D}(i_2,k_1)}\delta_{\mathfrak{D}(j_1,k_1),\mathfrak{D}(j_2,k_2)}\left(\overbar{\lambda}_{k_2}^{\mathfrak{B}(i_1)}\lambda_{k_1}^{'\ \mathfrak{B}(i_1)}(B_{k_2}^{\mathfrak{B}(i_1)})_{\mathfrak{C}(i_1),\mathfrak{D}(i_1,k_2)}(B_{k_1}^{\mathfrak{B}(i_1)})_{\mathfrak{C}(i_2),\mathfrak{D}(i_2,k_1)}\right.\nonumber\\
&\times&\left.\left.\overbar{\lambda}_{k_1}^{\mathfrak{B}(j_1)}\lambda_{k_2}^{'\ \mathfrak{B}(j_1)}(B_{k_1}^{\mathfrak{B}(j_1)})_{\mathfrak{D}(j_1,k_1),\mathfrak{C}(j_1)}B_{k_2}^{\mathfrak{B}(j_1)})_{\mathfrak{D}(j_2,k_2),\mathfrak{C}(j_2)}\right)\right).
\label{1d-ab}
\end{eqnarray}
The factor $\delta_{i_1,i_2}\delta_{j_1,j_2}$ comes from the argument used before, when the same $2D$-block matrices are matched in the bra and in the ket. The other Kr\"onecker symbols can be worked out explicitly, according to  $i_1$ and $i_2$ (resp.  $j_1$ and $j_2$) being equal or not, and by detailing whether the $2D$-block matrices are of $\mathcal{D}$ or $\overbar{\mathcal{D}}$-type, as was done in Eqs.(\ref{1b-ab}) and (\ref{1c-ab}). This is not done for Eq.(\ref{1d-ab}) to alleviate the presentation.\\

The formula for case (ii) can be deduced from Eq.(\ref{1d-ab}) by eliminating the terms where either $\overbar{\lambda}_{k_1}^{\mathfrak{B}(j_1)}$ or $\lambda_{k_1}^{'\ \mathfrak{B}(j_1)}$ appears, which leaves only the first term.
Similarly, for case (iii), one must eliminate the terms where either $\overbar{\lambda}_{k_1}^{\mathfrak{B}(i_1)}$ or $\lambda_{k_1}^{'\ \mathfrak{B}(i_1)}$ appears, which leaves only the second term on the second line. Care must be taken to count only once, the pairs $\{k_1,k_2\}$ such that one element is in $\overbar{\mathcal{U}}(\mathfrak{B}(i_1))\bigcap\mathcal{U}(\mathfrak{B}(j_1))$ and the other in $\mathcal{U}(\mathfrak{B}(i_1))\bigcap\overbar{\mathcal{U}}(\mathfrak{B}(j_1))$, which appear both in case (ii) and in case (iii).\\

\underline{\textbf{Case 2: $\mathfrak{B}(i_1)=\mathfrak{B}(j_1)$}}\\
No broken block in the bra, then, there cannot be broken block in the ket too, so necessarily $\mathfrak{B}(i_2)=\mathfrak{B}(j_2)$.\\

\underline{Subcase 2a: $\mathfrak{B}(i_1)\ne\mathfrak{B}(i_2)$}\\
$\mathfrak{B}(i_1)$ has to appear within $x_1,\ldots,x_n$ but not $\mathfrak{B}(i_2)$ since it will disappear from the ket after annihilation of $\varphi_{i_2}\wedge \overbar{\varphi}_{j_2}$ in it. Symmetrically, $\mathfrak{B}(i_2)$ has to appear within $y_1,\ldots,y_n$ but not $\mathfrak{B}(i_1)$. The remaining $n-1$ blocks in each set must be two by two identical because of $1$-orthogonality between the blocks, and each pair must be associated with the same geminal because of the EPI2O conditions. This means also that $\mathfrak{B}(i_1)$ in the bra and $\mathfrak{B}(i_2)$ in the ket  must be associated with the same  geminal  of index, say $k$. This is only possible if $\mathcal{U}(\mathfrak{B}(i_1))\bigcap\mathcal{U}(\mathfrak{B}(i_2))\ne\emptyset$. When this is the case, we obtain for $2D$-blocks:
\begin{eqnarray}
^2\Gamma^{\alpha,\beta}_{i_1j_1,i_2j_2}=\delta_{i_1,j_1}\delta_{i_2,j_2}\sum\limits_{k\in\mathcal{D}(\mathfrak{B}(i_1))\bigcap\mathcal{D}(\mathfrak{B}(i_2))}\overbar{\lambda}_{k}^{\mathfrak{B}(i_1)}\lambda_{k}^{'\ \mathfrak{B}(i_2)}(B_{k}^{\mathfrak{B}(i_1)})_{\mathfrak{C}(i_1),\mathfrak{C}(i_1)}(B_{k}^{\mathfrak{B}(i_2)})_{\mathfrak{C}(i_2),\mathfrak{C}(i_2)}\mathcal{S}_{n\smallsetminus\{k\}}^{m'\smallsetminus\{\mathfrak{B}(i_1),\mathfrak{B}(i_2)\}}\nonumber\\
+(1-\delta_{i_1,j_1})\delta_{i_2,j_2}\sum\limits_{k\in\overbar{\mathcal{D}}(\mathfrak{B}(i_1))\bigcap\mathcal{D}(\mathfrak{B}(i_2))}\overbar{\lambda}_{k}^{\mathfrak{B}(i_1)}\lambda_{k}^{'\ \mathfrak{B}(i_2)}(B_{k}^{\mathfrak{B}(i_1)})_{\mathfrak{C}(i_1),\mathfrak{C}(j_1)}(B_{k}^{\mathfrak{B}(i_2)})_{\mathfrak{C}(i_2),\mathfrak{C}(i_2)}\mathcal{S}_{n\smallsetminus\{k\}}^{m'\smallsetminus\{\mathfrak{B}(i_1),\mathfrak{B}(i_2)\}}\nonumber\\
+\delta_{i_1,j_1}(1-\delta_{i_2,j_2})\sum\limits_{k\in\mathcal{D}(\mathfrak{B}(i_1))\bigcap\overbar{\mathcal{D}}(\mathfrak{B}(i_2))}\overbar{\lambda}_{k}^{\mathfrak{B}(i_1)}\lambda_{k}^{'\ \mathfrak{B}(i_2)}(B_{k}^{\mathfrak{B}(i_1)})_{\mathfrak{C}(i_1),\mathfrak{C}(i_1)}(B_{k}^{\mathfrak{B}(i_2)})_{\mathfrak{C}(i_2),\mathfrak{C}(j_2)}\mathcal{S}_{n\smallsetminus\{k\}}^{m'\smallsetminus\{\mathfrak{B}(i_1),\mathfrak{B}(i_2)\}}\nonumber\\
+(1-\delta_{i_1,j_1})(1-\delta_{i_2,j_2})\!\!\!\!\!\!\sum\limits_{k\in\overbar{\mathcal{D}}(\mathfrak{B}(i_1))\bigcap\overbar{\mathcal{D}}(\mathfrak{B}(i_2))}\overbar{\lambda}_{k}^{\mathfrak{B}(i_1)}\lambda_{k}^{'\ \mathfrak{B}(i_2)}(B_{k}^{\mathfrak{B}(i_1)})_{\mathfrak{C}(i_1),\mathfrak{C}(j_1)}(B_{k}^{\mathfrak{B}(i_2)})_{\mathfrak{C}(i_2),\mathfrak{C}(j_2)}\mathcal{S}_{n\smallsetminus\{k\}}^{m'\smallsetminus\{\mathfrak{B}(i_1),\mathfrak{B}(i_2)\}}.\nonumber\\
\label{2a}
\end{eqnarray}
or more compactly (but at the price of summing terms that are zero),
\begin{eqnarray}
^2\Gamma^{\alpha,\beta}_{i_1j_1,i_2j_2}=\sum\limits_{k\in\mathcal{U}(\mathfrak{B}(i_1))\bigcap\mathcal{U}(\mathfrak{B}(i_2))}\overbar{\lambda}_{k}^{\mathfrak{B}(i_1)}\lambda_{k}^{'\ \mathfrak{B}(i_2)}(B_{k}^{\mathfrak{B}(i_1)})_{\mathfrak{C}(i_1),\mathfrak{C}(j_1)}(B_{k}^{\mathfrak{B}(i_2)})_{\mathfrak{C}(i_2),\mathfrak{C}(j_2)}\mathcal{S}_{n\smallsetminus\{k\}}^{m'\smallsetminus\{\mathfrak{B}(i_1),\mathfrak{B}(i_2)\}}.\nonumber\\
\label{2a-compact}
\end{eqnarray}
It is possible for some blocks to be $1D$. Then, $\mathcal{U}(\mathfrak{B}(i_1))\bigcap\mathcal{U}(\mathfrak{B}(i_2))$ can contain at most one geminal index, say $k$. If $\mathfrak{B}(i_1)<0$ and $\mathfrak{B}(i_2)>0$ then necessarily $i_1=j_1$ and :
\begin{eqnarray}
^2\Gamma^{\alpha,\beta}_{i_1i_1,i_2j_2}=\overbar{\lambda}_{k}^{\mathfrak{B}(i_1)}\lambda_{k}^{'\ \mathfrak{B}(i_2)}(B_{k}^{\mathfrak{B}(i_2)})_{\mathfrak{C}(i_2),\mathfrak{C}(j_2)}\mathcal{S}_{n\smallsetminus\{k\}}^{m'\smallsetminus\{\mathfrak{B}(i_2)\}},
\label{2a-compact-1D1}
\end{eqnarray}
else if $\mathfrak{B}(i_2)<0$ and $\mathfrak{B}(i_1)>0$ then necessarily $i_2=j_2$ and:
\begin{eqnarray}
^2\Gamma^{\alpha,\beta}_{i_1j_1,i_2i_2}=\overbar{\lambda}_{k}^{\mathfrak{B}(i_1)}\lambda_{k}^{'\ \mathfrak{B}(i_2)}(B_{k}^{\mathfrak{B}(i_1)})_{\mathfrak{C}(i_1),\mathfrak{C}(j_1)}\mathcal{S}_{n\smallsetminus\{k\}}^{m'\smallsetminus\{\mathfrak{B}(i_1)\}}.
\label{2a-compact-1D2}
\end{eqnarray}
For both $\mathfrak{B}(i_1)<0$ and $\mathfrak{B}(i_2)<0$ to occur, it is necessary that $h_k-h_{k-1}>1$ as   $\mathfrak{B}(i_1),\mathfrak{B}(i_2)\in\{h_{k-1}+1,\ldots,h_k\}$, then
\begin{eqnarray}
^2\Gamma^{\alpha,\beta}_{i_1i_1,i_2i_2}=\overbar{\lambda}_{k}^{\mathfrak{B}(i_1)}\lambda_{k}^{'\ \mathfrak{B}(i_2)}\mathcal{S}_{n\smallsetminus\{k\}}^{m'}.\nonumber
\label{2a-compact-1D12}
\end{eqnarray}

\underline{Subcase 2b: $\mathfrak{B}(i_1)=\mathfrak{B}(i_2)$}\\
$\mathfrak{B}(i_1)$ has to appear within $x_1,\ldots,x_n$ and  within $y_1,\ldots,y_n$. The remaining $n-1$ blocks in each set must be two by two identical because of $1$-orthogonality between the blocks, and each pair must be associated with the same geminal because of the EPI2O conditions. This means also that $\mathfrak{B}(i_1)$ in the bra and in the ket  must be associated with the same  geminal  of index, say $k\in\mathcal{U}(\mathfrak{B}(i_1))$.  When this is the case, we obtain for $2D$-blocks:
\begin{eqnarray}
^2\Gamma^{\alpha,\beta}_{i_1j_1,i_2j_2}=\sum\limits_{k\in\mathcal{U}(\mathfrak{B}(i_1))}\overbar{\lambda}_{k}^{\mathfrak{B}(i_1)}\lambda_{k}^{'\ \mathfrak{B}(i_1)}(B_{k}^{\mathfrak{B}(i_1)})_{\mathfrak{C}(i_1),\mathfrak{C}(j_1)}(B_{k}^{\mathfrak{B}(i_1)})_{\mathfrak{C}(i_2),\mathfrak{C}(j_2)}\mathcal{S}_{n\smallsetminus\{k\}}^{m'\smallsetminus\{\mathfrak{B}(i_1)\}}.\nonumber
\label{2b-compact}
\end{eqnarray}
If $\mathfrak{B}(i_1)<0$ then necessarily $i_1=j_1=i_2=j_2$ and let $k$ be the unique element of $\mathcal{U}(\mathfrak{B}(i_1))$:
\begin{eqnarray}
^2\Gamma^{\alpha,\beta}_{i_1i_1,i_1i_1}=\overbar{\lambda}_{k}^{\mathfrak{B}(i_1)}\lambda_{k}^{'\ \mathfrak{B}(i_1)}\mathcal{S}_{n\smallsetminus\{k\}}^{m'}.\nonumber
\label{2b-compact-1D}
\end{eqnarray}

\vspace{0.8cm}

Let us now turn to $^2\Gamma^{\alpha,\alpha}_{i_1j_1,i_2j_2}$, the case of  $^2\Gamma^{\beta,\beta}_{i_1j_1,i_2j_2}$ being analogous.
Here, the Pauli principle makes it sufficient to consider indices $i_1,j_1,i_2,j_2\in\{1,\ldots,m\}$ such that $i_1<j_1$ and $i_2<j_2$:
\begin{equation*}
^2\Gamma^{\alpha,\alpha}_{i_1j_1,i_2j_2}=  \sum\limits_{\substack{-h_n\leq x_1,\ldots,x_n\leq m'\\
\text{all distinct}\\ 
-h_n\leq y_1,\ldots,y_n\leq m'\\
\text{all distinct}}}
\!\!\!\!\!\!\langle \varphi_{i_1}\wedge\varphi_{j_1}\dlsh \Phi_{1}^{x_1}\wedge\cdots\wedge\Phi_{n}^{x_n}| \varphi_{i_2}\wedge \varphi_{j_2}\dlsh\Phi_{1}^{'\ y_1}\wedge\cdots\wedge\Phi_{n}^{'\ y_n} \rangle.
\end{equation*}
After annihilation of the $\alpha$-spin-orbitals, the lone  $\beta$-spin-orbitals in the bra and in the ket must be paired, this imposes $\mathfrak{B}(i_1)=\mathfrak{B}(i_2)$ and $\mathfrak{B}(j_1)=\mathfrak{B}(j_2)$.\\

\underline{\textbf{Case 1: $\mathfrak{B}(i_1)\ne\mathfrak{B}(j_1)$}}\\
Two distinct blocks are broken and must be associated with the same pair of geminals in the bra and in the ket,  of indices say $(k_1,k_2)$. The two blocks being $1$-orthogonal, the algebra is the same whether the spins of the orbitals in each block are the same or not. The formulas for $^2\Gamma^{\alpha,\alpha}$ are the same as those for $^2\Gamma^{\alpha,\beta}$. However, only the cases where $i_1<j_1$ and $i_2<j_2$ are relevant.\\

\underline{\textbf{Case 2: $\mathfrak{B}(i_1)=\mathfrak{B}(j_1)$}}\\
It is not possible to annihilate two spin-orbitals of the same spin in a given block, since all the blocks only contains $(S_z=0)$-geminals. So this case does not give any non-zero contribution.\\

The expressions obtained for $^2\Gamma$ are extremely simple to compute, provided that we have an efficient way of calculating the $\mathcal{S}_{n\smallsetminus\{u_1,\ldots,u_p\}}^{m'\smallsetminus\{y_1,\ldots,y_q\}}$'s with $p,q\le 2$.\\

\subsection{Expression of gradients for the EPI2O 2D-block geminal ansatz}
In the  EPI2O 2D-block geminal ansatz, there are three types of continuous
parameters:\\
(i) The coefficients defining the orbitals $\phi_i$'s in a given basis set.
We will not attempt to optimize them in the present study.\\ (ii) The block coefficients
$\lambda_k^x$'s.\\ 
(iii) In the case of $3$-type-$2D$-blocks, the angles $\theta_x$'s.\\
In order to optimize the last two kinds, we provide formulas to compute the gradients
with respect to their variations. We will assume that the $\lambda_k^x$'s are real,
so that $\overbar{\lambda}_k^x=\lambda_k^x$.

\subsubsection{Gradient with respect to block coefficients}
We first note that all the formulas of the previous section for the $2RTM$, hence for the $2RDM$,
are of the form:
\begin{equation}
^2\Gamma^{\gamma_1,\gamma_2}_{i_1j_1,i_2j_2}= \sum_w \mathbf{B}_w\Lambda_w\mathcal{S}_{n\smallsetminus K_w}^{m'\smallsetminus X_w},
\label{2RDM-el}
\end{equation}
where $\mathbf{B}_w$ is a product of block matrix elements (reduced to $1$ in the case of $1D$-block),
which can depend upon $\theta_x$'s but not upon $\lambda_k^x$'s; $\Lambda_w$ is a product of block matrix coefficients,
$\lambda_k^x$'s, which does not depend upon $\theta_x$'s; $K_w$ is a set of at most two $2D$-block indices
and $X_w$ is a set of at most two geminal indices. Like $\Lambda_w$, the quantity $\mathcal{S}_{n\smallsetminus K_w}^{m'\smallsetminus X_w}$
only depends upon the $\lambda_k^x$'s.\\

Taking the derivatives  with respect to $\lambda_k^x$ in  Eq.(\ref{2RDM-el}) gives:
\begin{equation}
\frac{\partial^2\Gamma^{\gamma_1,\gamma_2}_{i_1j_1,i_2j_2}}{\partial\lambda_k^x}= \sum_w \mathbf{B}_w\left(\frac{\partial\Lambda_w}{\partial\lambda_k^x}\mathcal{S}_{n\smallsetminus K_w}^{m'\smallsetminus X_w}+\Lambda_w\frac{\partial\mathcal{S}_{n\smallsetminus K_w}^{m'\smallsetminus X_w}}{\partial\lambda_k^x}\right).
\end{equation}
However, for a given $w$, at most one of the two terms can be non-zero: 
$\frac{\partial\Lambda_w}{\partial\lambda_k^x}$ will be non-zero only if $k\in K_w$,
whereas $\frac{\partial\mathcal{S}_{n\smallsetminus K_w}^{m'\smallsetminus X_w}}{\partial\lambda_k^x}$
will be non-zero only if $k\notin K_w$.\\

Let us consider $\frac{\partial\Lambda_w}{\partial\lambda_k^x}$:\\ 
In "case 1", it will be non-zero only if $k\in K_w$
and $x\in\mathfrak{B}(i_1)\bigcup\mathfrak{B}(j_1)$, then:
\begin{equation}
\frac{\partial\Lambda_w}{\partial\lambda_k^x}=\left\{\begin{array}{l l}
\frac{\Lambda_w}{\lambda_k^x} & \ \text{in the last two terms of case 1d(i)} \smallskip\\
\frac{2\Lambda_w}{\lambda_k^x} &  \text{in all other cases}\smallskip\\
 \end{array}\right. \quad.
\end{equation}

In "case 2", it will be non-zero only if $k\in K_w$
and $x\in\mathfrak{B}(i_1)\bigcup\mathfrak{B}(i_2)$, then:
\begin{equation}
\frac{\partial\Lambda_w}{\partial\lambda_k^x}=\left\{\begin{array}{l l}
\frac{\Lambda_w}{\lambda_k^x} & \ \text{if } \mathfrak{B}(i_1)\ne\mathfrak{B}(i_2)\smallskip\\
\frac{2\Lambda_w}{\lambda_k^x} &  \text{if } \mathfrak{B}(i_1)=\mathfrak{B}(i_2)\smallskip\\
 \end{array}\right. \quad.
\end{equation}

Regarding $\frac{\partial\mathcal{S}_{n\smallsetminus K_w}^{m'\smallsetminus X_w}}{\partial\lambda_k^x}$, 
it will be non-zero only if $k\notin K_w$, then:
\begin{equation}
\frac{\partial\mathcal{S}_{n\smallsetminus K_w}^{m'\smallsetminus X_w}}{\partial\lambda_k^x}=\left\{\begin{array}{l l}
2\lambda_k^x \mathcal{S}_{n\smallsetminus K_w\bigcup\{k\}}^{m'\smallsetminus X_w}& \ \text{if } x<0\smallskip\\
4\lambda_k^x \mathcal{S}_{n\smallsetminus K_w\bigcup\{k\}}^{m'\smallsetminus X_w\bigcup\{x\}}&  \text{if } x>0\smallskip\\
 \end{array}\right. \quad.
\end{equation}
Note that these gradients make sense only for coefficients $\lambda_k^x$ such that $k\in\mathcal{U}(x)$.

\subsubsection{Gradient with respect to angle parameters}
In the case of a $3$-type-$2D$-block of index $x$, $1\leq x\leq m''$, taking the derivatives  with respect to $\theta_x$ in  Eq.(\ref{2RDM-el}) gives:
\begin{equation}
\frac{\partial^2\Gamma^{\gamma_1,\gamma_2}_{i_1j_1,i_2j_2}}{\partial\theta_x}= \sum_w \frac{\partial\mathbf{B}_w}{\partial\theta_x}\Lambda_w\mathcal{S}_{n\smallsetminus K_w}^{m'\smallsetminus X_w}.
\end{equation}
The quantity $\frac{\partial\mathbf{B}_w}{\partial\theta_x}$ will be non-zero only if $k\in K_w$, and:\\
For case 1, if $B_k^x\in\mathcal{D}(\mathfrak{B}(i_1))\bigcup\mathcal{D}(\mathfrak{B}(j_1))$,
then 
\begin{equation}
\frac{\partial\mathbf{B}_w}{\partial\theta_x}=\left\{\begin{array}{l l}
 \text{in the last two terms of case 1d(i)}&\left\{\begin{array}{l l}
cot\theta_x \mathbf{B}_w &  \text{if }  (B_k^x)_{1,1} \text{ appears in } \mathbf{B}_w\smallskip\\
- tan\theta_x \mathbf{B}_w &  \text{if }(B_k^x)_{2,2} \text{ appears in } \mathbf{B}_w\smallskip\\
\end{array}\right.\smallskip\\ 
\text{in all other cases} &\left\{\begin{array}{l l}
2\ cot\theta_x \mathbf{B}_w & \ \text{if } (B_k^x)^2_{1,1} \text{ appears in } \mathbf{B}_w\smallskip\\
-2\ tan\theta_x \mathbf{B}_w & \ \text{if } (B_k^x)^2_{2,2} \text{ appears in } \mathbf{B}_w\smallskip\\
 \end{array}\right.
 \end{array}\right. \quad.
\end{equation}
For case 2, if $B_k^x\in\mathcal{D}(\mathfrak{B}(i_1))\bigcup\mathcal{D}(\mathfrak{B}(i_2))$,
then, 
\begin{equation}
\frac{\partial\mathbf{B}_w}{\partial\theta_x}=\left\{\begin{array}{l l}
  \text{if } \mathfrak{B}(i_1)\ne\mathfrak{B}(i_2)&\left\{\begin{array}{l l}
 cot\theta_x \mathbf{B}_w&  \text{if }  (B_k^x)_{1,1} \text{ appears in } \mathbf{B}_w\smallskip\\
- tan\theta_x \mathbf{B}_w&  \text{if }(B_k^x)_{2,2} \text{ appears in } \mathbf{B}_w\smallskip\\
\end{array}\right.\smallskip\\ 
 \text{if } \mathfrak{B}(i_1)=\mathfrak{B}(i_2) &\left\{\begin{array}{l l}
2\ cot\theta_x \mathbf{B}_w& \ \text{if } (B_k^x)^2_{1,1} \text{ appears in } \mathbf{B}_w\smallskip\\
 -2\ tan\theta_x \mathbf{B}_w& \ \text{if } (B_k^x)^2_{2,2} \text{ appears in } \mathbf{B}_w\smallskip\\
(cot\theta_x -  tan\theta_x) \mathbf{B}_w& \ \text{if } (B_k^x)_{1,1}(B_k^x)_{2,2} \text{ appears in } \mathbf{B}_w\smallskip\\
 \end{array}\right.
 \end{array}\right. \quad.
\end{equation}

\section{Proof of concept}
We have implemented our new geminal ansatz in the Nice-branch of the TONTO~\cite{Tonto} quantum chemistry code. We have already noted (see Remark 1) that two 1D-blocks with non-zero coefficients associated with the same geminal can always be replaced by a $3$-type-$2D$-block
with a $G_\theta$-submatrix attributed to that geminal, the optimization of the two 1D $\l$-coefficients amounting to that of the $\l$-coefficient and the $\theta$-angle of the $G_\theta$-submatrix. So, provided that the Hilbert subspaces associated with the geminals of an APSG wave function have all even dimensions, we can reproduce it  exactly by using $3$-type-$2D$-blocks and $G_\theta$-submatrices only. In the present study, we want to show within the simple frame of $3$-type-$2D$-blocks, that the accuracy of the new ansatz goes beyond that of APSG. Although we have implemented the general ansatz in TONTO, we choose to focus here on EPI2O-geminals made of $3$-type-$2D$-blocks only. It is sufficient to encompass APSG in many cases, and to go beyond it, as we shall see.\\

Moreover, we will limit our study to the singlet spin-restricted case. So, for each block, we will have a choice to allocate non-zero coefficients between only two types: a $G_\theta$-type or a $\s_x$-type. Assuming $m$ even, there will be $\frac{m}{2}$-blocks. For each of these blocks, the position of the $G_\theta$-type will be fixed to allow an APSG  approximate solution to be reproduced according to our previous remark. Then, it will remain to fix the position of the $\s_x$-types. There are $n-1$ positions left for each block, so,  in total, there are $(n-1)^{\frac{m}{2}}$ possibilities to allocate the $\s_x$ among the $n$ geminals.\\

The correlation energy gained with respect to the APSG model being due to $\s_x$-submatrices,  it comes exclusively from non-zero seniority configurations.  The difference between the full configuration interaction (FCI) energy of a molecular system and the lowest possible energy in the seniority-zero approximation, that is to say the doubly-occupied configuration interaction (DOCI) energy, being often maximum around the minimum of its Born-Oppenheimer potential energy surface (PES), (see for example Fig. (8a) of Ref. \cite{Fecteau2022}),  we will concentrate our study on molecular systems at their ground state equilibrium geometry.\\

To study the dependency upon the number of geminals with a fixed number of orbitals, we have chosen to study diatomics of the second row of the periodic table. A minimal basis set being too small to display significant dynamic correlation effects, we have used double zeta  with polarization bases, resulting in $28$ spherical orbitals, that is to say $14$ $2D$-blocks, for each system.  Actually, we also report results for a triple zeta basis, with $30$ $2D$-blocks to investigate basis set effects. However, the flexibility offered by such a large number of blocks was too large to be fully explored in this preliminary study. So, we just transposed the strategy for the allocation of block types that was derived from our double zeta examples. We have chosen to limit ourselves to homonuclear systems, so that respecting their symmetry helps  also to restrict the  possibilities of allocating  group types to the different geminals. The sample of selected molecules although small covers single, double and triple bound systems.\\

The results are presented in Tab. \ref{tab:results}. The difference between our best spin-restricted $3$-type-$2D$-block, so actually $2$-type-$2D$-block, $E^0_{\{G_\theta,\s_x\}}$  energy, and the APSG one, $E^0_{APSG}$, is of course due to the release of the strong orthogonality constraint, but can also be seen as resulting from the release of the seniority-zero constraint. So, it provides a lower bound to the correlation energy due to seniority non-zero configurations. We find this difference significant since it is more than the hundredth of Hartree for all the systems displayed. We cannot claim that the $E^0_{\{G_\theta,\s_x\}}$ energies are the lowest that can be obtained within our model, for we have not investigated all possible allocations of the $\s_x$-type non-zero coefficients among the geminals, neither have we tested  all the permutations of the molecular orbitals to form the 2D blocks. However, these results are sufficient to establish a proof of concept that the EPI2O model can be significantly more accurate than the APSG one.\\

Let us explain the strategy we have followed to obtain the results of  Tab. \ref{tab:results}. We start from APSG calculations. The optimized APSG geminals are numbered according to their associated energies in increasing order. So, geminals $1$ and $2$ are always degenerate core geminals, and the others with  higher energies, numbered from $3$ to $7$ in the case of N$_2$, from $3$ to $8$ for O$_2$, and from $3$ to $9$ for F$_2$, are valence geminals. The geminals are very much localized and are further classified into $3$ sets according to the partial charges associated with the two atoms. The group  where, say atom A, has partial charge $\approx 2$ and atom B $\approx 0$, is labelled "$2/0$", group "$0/2$" corresponds to the reverse, and group "$1/1$" to a partial charge of $\approx 1$ on each atom. For N$_2$, geminals $1$ and $4$  belong to group $2/0$ while their degenerate partners, geminals $2$ and $5$ to group $0/2$, the three geminals $3$, $6$ and $7$ to group $1/1$. A set of natural orbitals is associated with each $1$-orthogonal geminal and we rely on their populations to pair them into $2D$-blocks. That is to say, when the number of natural orbitals attached to a geminal is even, say $2k$, we make $k$ $2D$-blocks ordered according to decreasing absolute values of geminal expansion coefficients. When the number is odd, we transfer the orbital having the expansion coefficient with the least absolute value, to another geminal of the same partial charge group when possible. We try also  to only apply permutations that preserve geminal degeneracies. For example, the dimensions of the Hilbert subspaces of the geminals in the N$_2$ cc-pVTZ case, are:   $2\   2\  10\  14\  14\   9\   9$.
So, block $1$ is made of orbitals $1$ and $2$ associated with a $G_\theta$-type for geminal $1$,  block $2$ is made of orbitals $3$ and $4$ associated with a $G_\theta$-type for geminal $2$, blocks $3$ to $7$ are made of orbitals $5$ to $14$  all associated with  $G_\theta$-types for geminal $3$. However, since degenerate geminals $6$ and $7$ have odd numbers of natural orbitals,
we moved orbitals $51$ and $60$, which are their least occupied orbitals, to positions $15$ and $16$ respectively, to make a sixth $2D$-block associated with a $G_\theta$-type for geminal $3$, the latter being in the same $1/1$ group as geminals $6$ and $7$. Then, the next $7$, ($7,\ 4,\ 4$  respectively) $2D$-blocks are associated with  $G_\theta$-types for geminals $4$, ($5,\ 6,\ 7,$ respectively). The reordered geminal's natural orbitals together with the submatrix-type allocation for all the calculations presented in this article, can be found in input and output files available in the "tonto/tests/geminal" github repository. The allocation of the $\s_x$-types is done in the following way. For cc-pVDZ N$_2$ and O$_2$ cases, we found that only the $2D$-blocks associated with the natural orbitals with the highest occupancy in each geminal are improving the energy when a non-zero coefficient is attributed to a $\s_x$-type. This is why the number of parameters actually used has a value equal to $m+n$ in Tab. \ref{tab:results}, and not the maximum of $\frac{3m}{2}$ for the $2$-type model. For example, for the N$_2$ cc-pVDZ case, we used only $28+7=35$ parameters out of $42$. We find that the best energy is obtained by allocating the $\s_x$-types to geminals of the same group. In the N$_2$ example: we allocate a $\s_x$-type to geminal $4$ for the dominant block of geminal $1$ and conversely,  to geminal $5$ for the dominant block of geminal $2$ and conversely, to geminal $7$ for the dominant blocks of geminals $3$ and $6$, and to geminal $6$ for the dominant block of geminal $7$. In contrast, for F$_2$ one can gain $641$ microHartrees by using all of the $42$ available parameters.  This seems system-dependent rather than basis-dependent, since for N$_2$ in the cc-pVTZ basis, the energy obtained by using only $\s_x$-types for dominant blocks, that is to say $67$ parameters, is just $9$ microHartrees above the lowest energy we have managed to find for this system. The latter can be obtained with only $70$ parameters out of $90$. Usually, several different allocations are able to give the same energy and they may correspond to different numbers of terms in the overlap formula. The allocations with non-zero coefficients evenly spread among the geminals have more terms than those where the non-zero terms are packed in one or a few geminals. The "nb. terms in overlap" and CPU times displayed in Tab. \ref{tab:results}, correspond to allocations minimizing these two correlated quantities.
Note that the CPU times provided in the last line of Tab. \ref{tab:results} are only indicative of the order of magnitude, as no statistical processing has been performed.

\begin{table}[!ht]
    \centering
\begin{tabular}{|c|cccc|}
\multicolumn{5}{c}{\bf{Ground state energies from $2D$-block, $\{G_\theta,\s_x\}$-type calculations}}\\
\multicolumn{5}{c}{\bf{at  geometries derived from experiment (in  Hartree) }}\\
\multicolumn{5}{c}{\bf{}}\\
\hline
Homonuclear diatomics          & N$_2$      & N$_2$       & O$_2$        &  F$_2$      \\
\hline                                   
$r_e^{exp}$ ($\mathring{A}$) \footnotemark[1]   & $1.098$  & $1.098$ & $1.2075$ & $1.412$\\
Basis set \footnotemark[2]& cc-pVTZ & cc-pVDZ & cc-pVDZ  & cc-pVDZ\\
$E^0_{\{G_\theta,\s_x\}}$ \footnotemark[3] & -109.134390 & -109.082362 & -149.718718 & -198.853436\\
$E^0_{APSG}$ \footnotemark[4]  & -109.110630 & -109.058601 & -149.689436  & -198.840387 \\
$E^0_{RHF}$  & -108.983412 & -108.954087 & -149.542930 & -198.685664\\
nb. of geminals   & 7 & 7 & 8 &  9\\
nb. of blocks   & 30 & 14 & 14 &  14\\
nb. of parameters   & $70/90$ & $35/42$ & $36/42$ &  $42/42$\\
nb. terms in overlap  \footnotemark[5]      &  64974 &  360 & 343  & 1116\\
cpu time (s)\footnotemark[6]    & 0.4 & 0.01 & 0.01 &  0.02\\
\hline
\end{tabular}
\caption{Ground state energies (in  Hartree) from $2D$-block, $\{G_\theta,\s_x\}$-type calculations at  internuclear distances, $r_e^{exp}$  (in $\mathring{A}$), derived from experiment, for second row homonuclear diatomics. The $2D$-block, $\{G_\theta,\s_x\}$-type energies $E^0_{\{G_\theta,\s_x\}}$, can only be considered as upper bounds since the orbital grouping into $2D$-blocks  and the $\s_x$-type allocation to the different geminals have not been systematically investigated (see main text for details). Inputs and output files corresponding to these calculations are available in the "tests/geminal" sub-directory of the "tonto/Nice-branch" repository.  The line "nb. of parameters" gives the number of parameters actually used over the maximum possible within the model.}
\label{tab:results}
\footnotetext[1]{From NIST website: \url{https://cccbdb.nist.gov/}.}
\footnotetext[2]{From: T. H. Dunning, \jcp{90}{1007}{1989}}
\footnotetext[3]{Orbital grouping  and $\s_x$-type allocation have not been systematically investigated.}
\footnotetext[4]{From the SSG code of V. A. Rassolov in QCHEM: Y. Shao et al., \molphys{113}{184}{2015}}
\footnotetext[5]{For the block types allocation that gives the $E^0_{\{G_\theta,\s_x\}}$ energy reported.}
\footnotetext[6]{Sequential calculations, on a laptop with Intel® Core™ i9-10885H CPU $@$ 2.40GHz.   }

\end{table}

\section{Conclusion}

After an overview of existing APG-based methods, we have derived a general formula for the overlap between APG wave functions. It exhibits the full combinatorics complexity of APG-models and helps to understand the simplifications arising in known APG-constrained model. For example, the sum over the partitions of the number of geminals, $n$, reduces to a single term when the $1$-orthogonality constraint is enforced: the term with only $1$-cycle traces. In the model studied in the previous section, we have gone one step further: imposing the constraint that at most two coefficients per $2D$-block can be non-zero, implies that the sum over the partitions of $n$ are limited to the $([\frac{n}{2}]+1)$  terms having at most $2$-cycle traces, where $[k]$ denotes the integer part of $k$. More generally, in the case of spin-unrestricted $3$-type $2D$ blocks or spin-restricted $4$-type $2D$ blocks, there can be at most $3$ non-zero coefficients per block, therefore the partitions of $n$ are limited to terms involving at most $3$-cycle traces.  In the same way, in the case of spin-unrestricted $4$-type $2D$ blocks the sum on partitions of $n$ is limited to terms having at most $4$-cycle traces.\\

The permutationally invariant $2$-orthogonality constraints, proposed in Section \ref{PI2O},  restrict the double summation over permutations in the overlap formula. Considering again the spin-restricted $3$-type $2D$ blocks of the previous section, the $2$-orthogonality constraint (upper relation in Eq.(\ref{perm-2-orth})) implies that $\s=\s'$ on the geminal indices in $1$-cycle traces in Eq.(\ref{ovgen1}). The second relation allows one to restrict the $2$-cycle traces to those having $C$-matrices indices in increasing order. More precisely, let $\text{tr}\Big[C_{\sigma(i)}^{\dagger}C_{\sigma'(i)}C_{\sigma(j)}^{\dagger}C_{\sigma'(j)}\Big]$ be a general $2$-cycle trace, its invariance by circular permutation permits to choose $\sigma(i)<\sigma(j)$ without loss of generality, and, if $\sigma'(i)$, $\sigma(j)$ and $\sigma'(j)$ are all distinct,  the constraint $C_{\sigma'(i)}C_{\sigma(j)}^{\dagger}C_{\sigma'(j)}=-C_{\sigma'(j)}C_{\sigma(j)}^{\dagger}C_{\sigma'(i)}$ permits to choose $\sigma'(i)<\sigma'(j)$ without loss of generality, provided one keeps track of the sign factor.  So, the combination of the restriction on the number of non-zero coefficients per block, together with the permutationally invariant $2$-orthogonality (PI2O) conditions, keeps under control the combinatorial growth of the number of terms appearing in the overlap formula. This constitutes the extented PI2O (EPI2O) 2D-block geminal model, denoted as EPI2O-APG.\\

 We have obtained explicit formulas for the quantities necessary to implement the EPI2O-APG ansatz and provided a proof of principle that the latter model is able to give strictly lower energies than the APSG one. This has been accomplished in a simple version of the EPI2O-APG ansatz where the factors breaking the strong orthogonality constraint were all of  seniority greater than $0$. So, the extra correlation energy with respect to APSG was solely due to the removal of the seniority-zero constraint. The latter has been enforced in all recently proposed geminal approaches. The present study appears to be the first affordable geminal method taking into account seniority non-zero contributions. Although these contributions do not dominate quantitatively the energy, there are usually not homogeneous across a PES. So, our model could improve such properties as  dissociation energies, energy barriers or harmonic frequencies. This will be investigated in future works, together with an in-depth assessment of the computational performances of the EPI2O geminal ansatz.\\

The present work raises many questions and calls for many further developments. What are the best orbital block partition and the best submatrix-type allocation to geminals? The number of possibilities is huge. A number of empirical rules have been postulated in this study and partially confirmed. However, to deal with this flexibility, at least partially, it would still be desirable to optimize the molecular orbitals, so as to automatically create the best orbital block partition for a given submatrix-type allocation. In addition,  according to the variational principle, this would necessarily improve the results. Based on the example of orbital optimization for AP1roG wave functions~\cite{Limacher2014-mp,Boguslawski2014-jctc}, we can expect this improvement to be non negligible. The introduction of $4$-type $2D$-blocks should also be investigated, as this would add non strongly orthogonal components of seniority zero in the wave function, not present in the simplified model we have applied in the present article. So, this would partially separate out non $1$-orthogonal contributions from  non-zero seniority ones. Another direction for future developments is the implementation of the ansatz on hybrid quantum-classical computers. The structure of the wave function in terms of Pauli matrices (or $2\times 2$-diagonal matrices) hints to its encoding by qubits associated with pairs of spin-orbitals, as proposed in \cite{Khamoshi20,Elfving21}. In contrast with these works, the division by two of the number of qubits necessary for the wave function encoding would not be possible due to the release of the seniority-zero constraint. However, there could be a quantum advantage in calculating the energy for an EPI2O-APG wave function within a VQE-type algorithm \cite{Peruzzo14}. A quantum walk search algorithm could also be considered to explore the different submatrix-type allocations.\\

\section*{Acknowledgements}
This work has been supported by the French government, through the UCAJEDI Investments in the Future project managed by the National Research Agency (ANR) with the reference number ANR-15-IDEX-01, and the grant CARMA ANR-12-BS01-0017. Prof. V. Rassolov is acknowledged for fruitful discussions and his help in calculating $1$-orthogonal starting guesses. We are also thankful to J.-M. Lacroix and R. Ruelle for their support in using the cluster of the Lab. Dieudonné \\

\section{Appendix}
\subsection{Generalized seniority  \label{sen}}

The concept relies on a given decomposition of the one-electron Hilbert space, $\mathcal{H}$, into a direct sum of Hilbert subspaces called ``shells'', $\mathcal{H}=\mathcal{H}_1\oplus\mathcal{H}_2\oplus\cdots\oplus\mathcal{H}_n$, (where the dimension of $\mathcal{H}_i$ is $dim\mathcal{H}_i:=d_i$) or more conveniently, in practice, on a partition of the spin-orbital basis set into subsets, $(\chi_{i,j})_{j\in\{1,\ldots,d_i\}}$, spanning the shells of the decomposition, $i\in\{1,\ldots,n\}$ being the shell index~\cite{Perez2018}.\\

The generalized seniority number of a Slater determinant built over these spin-orbitals is simply the number of incomplete shells present in the determinant i.e. shells that are neither empty nor fully-occupied. More precisely, let $\chi_{i_1,j_1}\wedge\cdots\wedge\chi_{i_1,j_{k_1}}\wedge\cdots\wedge\chi_{i_p,j_p}\wedge\cdots\wedge\chi_{i_p,j_{k_p}}$ be such a Slater determinant (where $\wedge$ denotes the intrinsically antisymmetrical, ``Grassmann'' or ``exterior'' product \cite{Cassam2003-jmp}), one counts $1$ for every $k_m$ such that $0<k_m<d_m$ (assuming all spin-orbitals are distinct otherwise the determinant is zero). This definition extends over linear combinations of Slater determinants having same generalized seniority number. Then, note that the definition is invariant under arbitrary rotations (or more generally, unitary transformations) of the spin-orbitals within the same shell. Hence, it is legitimate to talk of seniority with respect to the shell decomposition and not just with respect to the spin-orbital partition.\\

When there are $m$ orbitals, one can retrieve the usual notion of seniority \cite{Racah1943,Dean2003,Zelevinsky2003,Bytautas2011} by decomposing the Hilbert space into $m$ two-dimensional Hilbert subspaces, $\mathcal{H}=\mathcal{H}_1\oplus\cdots\oplus\mathcal{H}_m$, where each $\mathcal{H}_i$ is spanned by two  spin-orbitals of spin-$\a$ and  spin-$\b$, respectively, with the same spatial part. However, other partitions not only based on spin-degeneracy, can prove physically relevant, such as partitions into atomic shell orbitals (hence the name we have coined), or into spatially degenerate molecular orbitals.\\

\subsection{p-orthogonality \label{ortho}}
Let $p>0$ be an integer, $\Psi_1$  an $n_1$-electron wave function, and $\Psi_2$  an $n_2$-electron one, with $n_1\geq p$ and $n_2\geq p$. 
Let us consider, their $p$-particle reduced density matrices, $^p\Gamma_1$ and $^p\Gamma_2$ respectively. We call ``$p$-external space of 
$\Psi_i$'' the $p$-particle Hilbert subspace spanned by the unoccupied $p$-particle natural wave functions, that is to say, by the eigenfunctions of $^p\Gamma_i$ associated with the $0$ eigenvalue, for $i\in\{1,2\}$, and ``$p$-internal space of $\Psi_i$'', noted $\mathcal{I}[\Psi_i]$ the orthogonal subspace. In other words, $\mathcal{I}[\Psi_i]$  is the Hilbert space spanned by the partially occupied  $p$-particle natural wave functions. We will say that $\Psi_1$ and $\Psi_2$ are $p$-orthogonal if and only if their $p$-internal spaces are orthogonal: $\mathcal{I}[\Psi_1]\bot \mathcal{I}[\Psi_2]$. Note that  $1$-orthogonality is nothing but the so-called ``strong orthogonality'' \cite{Parr56,McWeeny59,McWeeny61}, and when $n_1=n_2=n$, $n$-orthogonality is just the usual concept of orthogonality. For any integers $0<p\leq q\leq n$, we have the important property that $p$-orthogonality implies $q$-orthogonality, which means that the lesser the integer $p$, the stronger the $p$-orthogonality. An earlier definition of this concept is due to S. Wilson \cite{Wilson76}.\\

\subsection{Proof of the general APG overlap formula \label{proof}}
We will prove formula (\ref{ovgen1}), by using the Hopf algebra formalism, which  simplifies the derivation of several famous results and suggests natural generalizations of the latter\cite{Cassam2003-jmp}. We believe that the so-called  "co-product" of the Hopf algebra of a Fermionic system, is the  most appropriate mathematical tool to tackle the proof, since it allows one to break an antisymmetrized product wave function into two parts while preserving antisymmetry. The combinatorial and sign factors related to the action of the Symmetric group are encapsulated in the formalism. 

We first rewrite the general APG overlap formula by introducing an alternative partition of the different factors:
\begin{equation} 
 \langle\Psi_e|\Psi'_e\rangle=\langle \Phi_1\wedge\cdots\wedge\Phi_n | \Phi'_1\wedge\cdots\wedge\Phi'_n \rangle=\sum_{\substack{0\leq N_{n,0},\ldots, N_{n,n}\leq n \\ \sum\limits_{k=0}^n N_{n,k}=\sum\limits_{k=0}^n k N_{n,k}=n}} Q_{N_{n,0},\ldots,N_{n,n}} \sum_{\sigma,\sigma'\in\mathfrak{S}_n} \prod_{k=1}^n  P_{N_{n,0},\ldots,N_{n,k-1}}^{N_{n,k}} (\sigma,\sigma')
 \label{ovgen3}
\end{equation}
with $Q_{N_{n,0},\ldots,N_{n,n}}$ gathering the normalization and sign factors:
\begin{equation}
 Q_{N_{n,0},\ldots,N_{n,n}} = \frac{(-1)^{N_{n,0}}}{\prod\limits_{k=1}^n k^{N_{n,k}}\, N_{n,k}!}\quad,
  \label{ovgen4}
\end{equation}
and 
\begin{equation}
 P_{N_{n,0},\ldots,N_{n,k-1}}^{N_{n,k}} (\sigma,\sigma') = \prod_{j=1}^{N_{n,k}}  T_{N_{n,0},\ldots,N_{n,k-1}}^{N_{n,k}} (j,\sigma,\sigma')\quad.
  \label{ovgen5}
\end{equation}
Our proof is obtained by mathematical induction. The case $n=1$ is trivial. Assuming that the equality is true for a particular $k\geq1$, we want to calculate the following scalar product:
\begin{equation*}
 \langle \Phi_1\wedge\cdots\wedge\Phi_{k+1} |  \Phi'_1\wedge\cdots\wedge\Phi'_{k+1} \rangle \quad.
\end{equation*}
In order to use the induction hypothesis, we transform the last exterior product in the bra into a right co-product, by Hopf duality. The left-hand side becomes:
\begin{equation*}
 \Phi_1\wedge\cdots\wedge\Phi_k\otimes\Phi_{k+1} \quad,
\end{equation*}
while the action of the co-product on the right-hand side gives the following expression:
\begin{align*}
 &\!\!\!\!\!\!\!\!\!\!\!\Phi'_1\wedge\cdots\wedge\Phi'_k\otimes\Phi'_{k+1}+\sum_{u=1}^k\Phi'_1\wedge\cdots\wedge\Phi'_{u-1}\wedge\Phi'_{k+1}\wedge\Phi'_{u+1}\wedge\cdots\wedge\Phi'_k\otimes\Phi'_u\\
 &\!\!-\sum_{1\leq\alpha,\beta,\gamma,\delta\leq m}\sum_{1\leq u<v\leq k+1}[(C'_u)_{\gamma,\beta}(C'_v)_{\alpha,\delta}+(C'_v)_{\gamma,\beta}(C'_u)_{\alpha,\delta}]\times\\
 &\quad\quad\times \Phi'_1\wedge\cdots\wedge\Phi'_{u-1}\wedge\Phi'_{u+1}\wedge\cdots\wedge\Phi'_{v-1}\wedge\Phi'_{v+1}\wedge\cdots\wedge\Phi'_{k+1}\wedge\varphi_{\alpha}\wedge\overbar{\varphi_{\beta}}\otimes(\varphi_{\gamma}\wedge\overbar{\varphi_{\delta}}) \quad.
\end{align*}
With this new formulation, we can develop each expression and apply the induction hypothesis. The scalar product can now be written as a sum of two distinct terms: $K_1$, composed by the terms in which the geminals of the right-hand side term have not been cut by the co-product, and $K_2$, formed by the other terms.\\
For $K_1$, we obtain:
\begin{align*}
 K_1 &= \langle \Phi_1\wedge\cdots\wedge\Phi_k |  \Phi'_1\wedge\cdots\wedge\Phi'_k \rangle\langle \Phi_{k+1} |  \Phi'_{k+1} \rangle + \sum_{u=1}^k\langle \Phi_1\wedge\cdots\wedge\Phi_k |  \bigwedge_{\substack{1\leq l\leq k+1 \\ l\neq u}}\Phi'_l \rangle\langle \Phi_{k+1} |  \Phi'_u \rangle\\
 &=\langle \Phi_1\wedge\cdots\wedge\Phi_k |  \Phi'_1\wedge\cdots\wedge\Phi'_k \rangle\,\text{tr}(C_{k+1}^\dagger C'_{k+1}) + \sum_{u=1}^k\langle \Phi_1\wedge\cdots\wedge\Phi_k |  \bigwedge_{\substack{1\leq l\leq k+1 \\ l\neq u}}\Phi'_l \rangle\,\text{tr}(C_{k+1}^\dagger C'_u) \quad.
\end{align*}
Applying the induction hypothesis on the two terms of $K_1$ (for the second one, by replacing $\Phi'_u$ with $\Phi'_{k+1}$ so $C'_u$ with $C'_{k+1}$), we can write:
\begin{equation}
 K_1 = \sum_{\substack{0\leq N_{k,0},\ldots, N_{k,k}\leq k \\ \sum\limits_{i=0}^k N_{k,i}=\sum\limits_{i=0}^k i N_{k,i}=k}} Q_{N_{k,0},\ldots,N_{k,k}} \sum_{u=1}^{k+1} \text{tr}(C_{k+1}^\dagger C'_u)\sum_{\sigma,\sigma'\in\mathfrak{S}_k} \prod_{i=1}^k P_{N_{k,0},\ldots,N_{k,i-1}}^{N_{k,i}} (\sigma,\tau_{u,k+1}\sigma')\quad,
 \label{K1}
\end{equation}
with $\tau_{u,k+1}$ the transposition $(u,k+1)\in\mathfrak{S}_{k+1}$.\\

For $K_2$ we have:
\begin{align*}
&\!\!\!\!\!\!\!\!\!\!\!\!\!\!\!\!\!\!\!\!\!\!\!\!\!\!\!\!\!\!\!\!K_2 = -\!\!\!\sum_{1\leq\alpha,\beta,\gamma,\delta\leq m}\sum_{1\leq u<v\leq k+1}\!\!\![(C'_u)_{\gamma,\beta}(C'_v)_{\alpha,\delta}+(C'_v)_{\gamma,\beta}(C'_u)_{\alpha,\delta}]\langle \Phi_1\wedge\cdots\wedge\Phi_k |  \Big(\bigwedge_{\substack{1\leq l\leq k+1 \\ l\neq u,v}}\Phi'_l\Big) \wedge\varphi_{\alpha}\wedge\overbar{\varphi_{\beta}} \rangle\langle \Phi_{k+1} |  \varphi_{\gamma}\wedge\overbar{\varphi_{\delta}} \rangle\\
 &\!\!\!\!\!\!\!\!\!\!\!\!\!\!\!\!\!\!\!\!\!\!\!\!= -\sum_{1\leq u<v\leq k+1}\sum_{1\leq\alpha,\beta\leq m}\Big[\langle \Phi_1\wedge\cdots\wedge\Phi_k |  \Big(\bigwedge_{\substack{1\leq l\leq k+1 \\ l\neq u,v}}\Phi'_l\Big) \wedge\varphi_{\alpha}\wedge\overbar{\varphi_{\beta}} \rangle\times \\
 &\quad\quad\quad\quad\quad\quad\quad\quad\quad\quad \times\sum_{1\leq\gamma,\delta\leq m}\Big((C'_v)_{\alpha,\delta}\langle \Phi_{k+1} |  \varphi_{\gamma}\wedge\overbar{\varphi_{\delta}} \rangle(C'_u)_{\gamma,\beta}+(C'_u)_{\alpha,\delta}\langle \Phi_{k+1} |  \varphi_{\gamma}\wedge\overbar{\varphi_{\delta}} \rangle(C'_v)_{\gamma,\beta}\Big)\Big]\\
 &\!\!\!\!\!\!\!\!\!\!\!\!\!\!\!\!\!\!\!\!\!\!\!\!= -\sum_{1\leq u<v\leq k+1}\sum_{1\leq\alpha,\beta\leq m}\langle \Phi_1\wedge\cdots\wedge\Phi_k |  \Big(\bigwedge_{\substack{1\leq l\leq k+1 \\ l\neq u,v}}\Phi'_l\Big) \wedge\varphi_{\alpha}\wedge\overbar{\varphi_{\beta}} \rangle \big[\big(C'_v C_{k+1}^\dagger C'_u\big)_{\alpha\beta}+\big(C'_u C_{k+1}^\dagger C'_v\big)_{\alpha\beta}\big]\quad.
\end{align*}
Let $E_{\alpha\beta}$ be the matrix where only the coefficient of line $\alpha$ and column $\beta$ is non-zero and equal to $1$. We can use the induction hypothesis by replacing $\Phi'_u$ with $\varphi_{\alpha}\wedge\overbar{\varphi_{\beta}}$ and, for $v\neq k+1$, $\Phi'_v$ with $\Phi'_{k+1}$ (i.e. by substituting $C'_u$ with $E_{\alpha\beta}$ and, for $v\neq k+1$, $C'_v$ with $C'_{k+1}$). To take advantage of the simple form of $E_{\alpha\beta}$, we have to rewrite our recursion formula as:
\begin{align*}
 \!\!\!\!\!\!\!\!\!\!\sum_{\substack{0\leq N_{k,0},\ldots, N_{k,k}\leq k \\ \sum\limits_{i=0}^k N_{k,i}=\sum\limits_{i=0}^k i N_{k,i}=k \\ N_{k,x_u}\geq 1}} &Q_{N_{k,0},\ldots,N_{k,k}} \sum_{x_u=1}^k N_{k,x_u}x_u \sum_{\substack{\sigma,\sigma'\in\mathfrak{S}_k \\ \sigma'(L_{x_u})=u}} \text{tr}\Big[\Big(\prod\limits_{l=L_{x_u-1}+(N_{k,x_u}-1)x_u+1}^{L_{x_u}-1} C_{\sigma(l)}^{\dagger}C'_{\sigma'(l)}\Big)C_{\sigma(L_{x_u})}^{\dagger}E_{\alpha\beta}\Big]\times\\
 &\times P_{N_{k,0},\ldots,N_{k,x_u-1}}^{N_{k,x_u}-1}(\sigma,\sigma')\prod_{\substack{1\leq i\leq k \\ i\neq x_u}}P_{N_{k,0},\ldots,N_{k,i-1}}^{N_{k,i}} (\sigma,\sigma')\quad,
\end{align*}
with $L_{x_u}=\sum\limits_{p=0}^{x_u} p N_{k,p}$.\\
Note that a multiplying factor has appeared. Indeed, since we have distinguished the trace containing $E_{\alpha\beta}$, our formula is no longer completely symmetric under permutations. We had to multiply by $N_{k,x_u}$, the number of possible positions of that specific $x_u$-cycle trace in its product with the other $x_u$-cycle traces (consequence of the commutativity of multiplication in $\mathbb{K}$), and then by $x_u$, which is the number of possible locations for $E_{\alpha\beta}$ in our trace (consequence of the invariance of the trace under circular permutations of matrices).\\
We also notice that $N_{k,x_u}$ cannot be zero since it is related to the trace containing $E_{\alpha\beta}$, which necessarily exists. This trace is actually reduced to a single matrix element:
\begin{equation}
 \!\!\!\!\!\!\!\!\!\!\!\!\!\text{tr}\Big[\Big(\prod\limits_{l=L_{x_u-1}+(N_{k,x_u}-1)x_u+1}^{L_{x_u}-1} C_{\sigma(l)}^{\dagger}C'_{\sigma'(l)}\Big)C_{\sigma(L_{x_u})}^{\dagger}E_{\alpha\beta}\Big] = \Big[\Big(\prod\limits_{l=L_{x_u-1}+(N_{k,x_u}-1)x_u+1}^{L_{x_u}-1} C_{\sigma(l)}^{\dagger}C'_{\sigma'(l)}\Big)C_{\sigma(L_{x_u})}^{\dagger}\Big]_{\beta\alpha}\quad.
\end{equation}
Then, a \textit{bona fide} trace  re-appears in  $K_2$ since sum:
\begin{equation*}
\sum_{1\leq\alpha,\beta\leq m} \big[\big(C'_v C_{k+1}^\dagger C'_u\big)_{\alpha\beta}+\big(C'_u C_{k+1}^\dagger C'_v\big)_{\alpha\beta}\big]\Big[\Big(\prod\limits_{l=L_{x_u-1}+(N_{k,x_u}-1)x_u+1}^{L_{x_u}-1} C_{\sigma(l)}^{\dagger}C'_{\sigma'(l)}\Big)C_{\sigma(L_{x_u})}^{\dagger}\Big]_{\beta\alpha}
\end{equation*}
is nothing but:
\begin{equation*}
 \text{tr}\Big[C_{\sigma(L_{x_u})}^{\dagger}\big(C'_v C_{k+1}^\dagger C'_u+C'_u C_{k+1}^\dagger C'_v\big)\prod\limits_{l=L_{x_u-1}+(N_{k,x_u}-1)x_u+1}^{L_{x_u}-1} C_{\sigma(l)}^{\dagger}C'_{\sigma'(l)}\Big] \quad.
\end{equation*}
By using the induction hypothesis, $K_2$ can be written:
\begin{align}
 &\!\!\!\!\!\!\!\!\!\!\!\!\!\!\!K_2=-\sum_{1\leq u<v\leq k+1}\sum_{x_u=1}^k\sum_{\substack{0\leq N_{k,0},\ldots, N_{k,k}\leq k \\ \sum\limits_{i=0}^k N_{k,i}=\sum\limits_{i=0}^k i N_{k,i}=k \\ N_{k,x_u}\geq 1}} Q_{N_{k,0},\ldots,N_{k,k}}N_{k,x_u}x_u \sum_{\substack{\sigma,\sigma'\in\mathfrak{S}_k \\ \sigma'(L_{x_u})=u}} P_{N_{k,0},\ldots,N_{k,x_u-1}}^{N_{k,x_u}-1}(\sigma,\tau_{v,k+1}\sigma')\times\nonumber\\
 &\!\!\!\!\!\!\!\!\!\!\!\!\!\!\!\!\!\!\!\!\!\!\!\!\!\!\!\times\text{tr}\Big[C_{\sigma(L_{x_u})}^{\dagger}\big(C'_v C_{k+1}^\dagger C'_u+C'_u C_{k+1}^\dagger C'_v\big)\prod\limits_{l=L_{x_u-1}+(N_{k,x_u}-1)x_u+1}^{L_{x_u}-1} C_{\sigma(l)}^{\dagger}C'_{\tau_{v,k+1}\sigma'(l)}\Big]\prod_{\substack{1\leq i\leq k \\ i\neq x_u}}P_{N_{k,0},\ldots,N_{k,i-1}}^{N_{k,i}}(\sigma,\tau_{v,k+1}\sigma')\quad,
  \label{K2}
\end{align}
with $\tau_{v,k+1}$ the transposition $(v,k+1)\in\mathfrak{S}_{k+1}$.\\

Note that the trace which is factored in the general term $K_2$ and whose length has increased from $2x_u$ to $2(x_u+1)$ actually represents two traces of $C$-matrix products since we have kept the symmetrized form $\big(C'_v C_{k+1}^\dagger C'_u+C'_u C_{k+1}^\dagger C'_v\big)$ inside in order to reduce the size of the  $K_2$-expression.\\

$K_1$ and $K_2$ have now a very similar form. However, we still have to replace the sum on partitions of $k$ with one on partitions of $k+1$. To perform this, we define several mappings.\\

First, let $\mathcal{F}_{k,0}$ denote the following mapping:
\begin{align*}
 \mathcal{F}_{k,0} : \quad \quad \quad \quad \ \mathbb{N}^{k+1} &\longrightarrow \mathbb{N}\times\mathbb{N}^*\times\mathbb{N}^{k-1}\times\{0\}\\
 (N_{k,0},\ldots,N_{k,k}) &\longmapsto (N_{k,0},N_{k,1}+1,N_{k,2},\ldots,N_{k,k},0) \quad.
\end{align*}
For $1\leq i\leq k-1$, we define $\mathcal{F}_{k,i}$ as:
\begin{align*}
 \mathcal{F}_{k,i} : \mathbb{N}^{i} \times \mathbb{N}^*\times \mathbb{N}^{k-i} &\longrightarrow \mathbb{N}^*\times\mathbb{N}^{i}\times\mathbb{N}^*\times\mathbb{N}^{k-i-1}\times\{0\}\\
 (N_{k,0},\ldots,N_{k,k}) &\longmapsto (N_{k,0}+1,N_{k,1},\ldots,N_{k,i-1},N_{k,i}-1,N_{k,i+1}+1,N_{k,i+2},\ldots,N_{k,k},0) \quad.
\end{align*}
Finally, we introduce $\mathcal{F}_{k,k}$ as follows:
\begin{align*}
 \mathcal{F}_{k,k} : \quad \quad \quad \mathbb{N}^{k} \times \mathbb{N}^* &\longrightarrow \mathbb{N}^*\times\mathbb{N}^{k}\times\{1\}\\
 (N_{k,0},\ldots,N_{k,k}) &\longmapsto (N_{k,0}+1,N_{k,1},\ldots,N_{k,k-1},N_{k,k}-1,1) \quad.
\end{align*}
These mappings transform a partition of $k$ into a partition of $k+1$. Moreover, they are invertible on their image and their inverses allow to map a partition of $k+1$ onto a partition of $k$:
\begin{align*}
 \mathcal{F}_{k,0}^{-1} : \quad \mathbb{N}\times\mathbb{N}^*\times\mathbb{N}^{k-1}\times\{0\} &\longrightarrow \mathbb{N}^{k+1}\\
 (N_{k+1,0},\ldots,N_{k+1,k},0) &\longmapsto (N_{k+1,0},N_{k+1,1}-1,N_{k+1,2},\ldots,N_{k+1,k}) \quad;
\end{align*}
\begin{align*}
 &\!\!\!\!\!\!\!\!\!\!\!\!\!\!\!\!\!\!\!\!\!\!\!\!\!\!\!\!\!\!\!\!\!\mathcal{F}_{k,i}^{-1} : \ \mathbb{N}^*\times\mathbb{N}^{i}\times\mathbb{N}^*\times\mathbb{N}^{k-i-1}\times\{0\} \longrightarrow \mathbb{N}^{i} \times \mathbb{N}^*\times \mathbb{N}^{k-i} \quad \quad (1\leq i\leq k-1)\\
 &\!\!\!\!\!\!\!\!\!\!\!\!\!\!\!\!\!\!(N_{k+1,0},\ldots,N_{k+1,k},0) \longmapsto (N_{k+1,0}-1,N_{k+1,1},\ldots,N_{k+1,i-1},N_{k+1,i}+1,N_{k+1,i+1}-1,N_{k+1,i+2},\ldots,N_{k+1,k}) \quad;
\end{align*}
\begin{align*}
 \mathcal{F}_{k,k}^{-1} : \quad \quad \quad \quad \mathbb{N}^*\times\mathbb{N}^{k}\times\{1\} &\longrightarrow \mathbb{N}^{k} \times \mathbb{N}^*\\
 (N_{k+1,0},\ldots,N_{k+1,k},1) &\longmapsto (N_{k+1,0}-1,N_{k+1,1},\ldots,N_{k+1,k-1},N_{k+1,k}+1) \quad.
\end{align*}

If we consider a partition of $k$ which has $q$ non-zero $N_{k,i}$'s ($i\geq 1$), then we can apply exactly $q+1$ distinct mappings $\mathcal{F}_{k,i}$ (one for each non-zero $N_{k,i}$, and $\mathcal{F}_{k,0}$ is always applicable), each giving a different partition of $k+1$.\\
Conversely, 
let $(N_{k+1,0},\ldots,N_{k+1,k+1})$ be an arbitrary partition of $k+1$. Consider $q\geq1$ and $i_1<\ldots<i_q$ such as the $N_{k+1,i_j+1}$'s are the non-zero integers determining our partition of $k+1$ (we deduce the value $N_{k+1,0}$). This partition can then stem from exactly $q$ different partitions of $k$, denoted by $(N_{k,0}^{i_j},\ldots,N_{k,k}^{i_j})$:
\begin{equation*}
 \!\!\!\!\!\!\!\!\!\!\!\!\!\!\!\!\!\!\!\!\!\!\!\!\!\left\{ \begin{array}{l l}
  &(N_{k,0}^{i_1},\ldots,N_{k,k}^{i_1}) = \mathcal{F}_{k,i_1}^{-1}(N_{k+1,0},\ldots,N_{k+1,k+1}) \medskip\\ 
  &(N_{k,0}^{i_2},\ldots,N_{k,k}^{i_2}) = \mathcal{F}_{k,i_2}^{-1}(N_{k+1,0},\ldots,N_{k+1,k+1}) \medskip\\
  &\quad\quad\quad\quad\quad\quad\quad\quad\quad\cdot \\
  &\quad\quad\quad\quad\quad\quad\quad\quad\quad\cdot \\
  &\quad\quad\quad\quad\quad\quad\quad\quad\quad\cdot \\
  &(N_{k,0}^{i_q},\ldots,N_{k,k}^{i_q}) = \mathcal{F}_{k,i_q}^{-1}(N_{k+1,0},\ldots,N_{k+1,k+1})
 \end{array} \right. \quad.
\end{equation*}
So, we have three possible forms for the preimages of $(N_{k+1,0},\ldots,N_{k+1,k+1})$.\\
(i) If $i_1=0$, then: 
\begin{equation}
 (N_{k,0}^0,\ldots,N_{k,k}^0) = (N_{k+1,0},N_{k+1,1}-1,N_{k+1,2},\ldots,N_{k+1,k}) \quad.
\end{equation}
We see from Eq.(\ref{K1}), that such a pre-image will arise from $K_1$.\\
(ii) If $i_q=k$, then $q=1$ and:
\begin{equation}
 (N_{k,0}^k,\ldots,N_{k,k}^k) = (N_{k+1,0}-1,N_{k+1,1},\ldots,N_{k+1,k-1},N_{k+1,k}+1) \quad.
\end{equation}
(iii) Finally, for the general case $1\leq i_j\leq k-1$:
\begin{equation}
 \!\!\!\!\!\!\!\!\!\!\!\!(N_{k,0}^{i_j},\ldots,N_{k,k}^{i_j}) = (N_{k+1,0}-1,N_{k+1,1},\ldots,N_{k+1,i_j-1},N_{k+1,i_j}+1,N_{k+1,i_j+1}-1,N_{k+1,i_j+2},\ldots,N_{k+1,k})\quad.
\end{equation}
We see from Eq.(\ref{K2}), that these last two cases will arise from the terms in $K_2$.\\

So, all the products of traces required for our partition of $k+1$, $(N_{k+1,0},\ldots,N_{k+1,k+1})$, will appear either in $K_1$ or in $K_2$, $C_{k+1}^\dagger$ being inserted in the trace that has been factored out from the product of traces, and the transpositions insuring that $C'_{k+1}$ will be inserted in all possible factors of the remaining product of traces. Now, we just have to examine the coefficients coming from the different pre-images.
Let us denote by $Q_{i_j}$ the factor coming from the $(N_{k,0}^{i_j},\ldots,N_{k,k}^{i_j})$ pre-image. The products of traces related to the latter are those, where $C_{k+1}^\dagger$ appears in a $(i_j+1)$-cycle trace. We  have to show that by transforming the summations in $K_1$ or $K_2$ into sums over $\mathfrak{S}_{k+1}$, these factors are all equal to the expected normalization factor: $Q_{N_{k+1,0},\ldots,N_{k+1,k+1}}$.\\

Let us first consider $Q_{0}$ (only present if $i_1=0$). The terms for which $Q_0$ is factored come from $K_1$. To find the value of $Q_0$ from $Q_{N_{k,0}^0,\ldots,N_{k,k}^0}$, we have to examine how the summations over $\sigma$, $\sigma'$ and $u$ can be compacted into a double summation over $\mathfrak{S}_{k+1}$.  To each $\sigma'\in\mathfrak{S}_k$, corresponds $k+1$ distinct  permutations $\sigma'_{k+1,u}\in\mathfrak{S}_{k+1}$:
\begin{equation}
 \left\{ \begin{array}{l l}
  \sigma'_{k+1,u}(\sigma'^{-1}(u))=k+1\\
  \sigma'_{k+1,u}(k+1)=u\\
  \sigma'_{k+1,u}(l)=\sigma'(l) \ \text{for all } l\in\{1,\ldots,k\} \setminus \{\sigma'^{-1}(u)\}
 \end{array} \right. \quad.
\end{equation}
In this way, the summation over $u$ from $1$ to $k+1$ and the one over $\sigma'\in\mathfrak{S}_k$ can be condensed into a sum over $\mathfrak{S}_{k+1}$ without changing the normalization factor already present.
To extend the summation over $\sigma$ from $\mathfrak{S}_k$ to $\mathfrak{S}_{k+1}$, we have to divide by the number of possibilities to interleave the new  $1$-cycle trace (the one containing $C_{k+1}^\dagger$) within the $(N_{k+1,1}-1)$ other traces, (that is to say  by $(N_{k+1,1}-1)+1$). So, we obtain:
\begin{align*}
 Q_0 &=  \underbrace{Q_{N_{k+1,0},N_{k+1,1}-1,N_{k+1,2},\ldots,N_{k+1,k}}}_{\text{factor $Q_{N_{k,0}^0,\ldots,N_{k,k}^0}$ appearing in $K_1$}}\times\underbrace{\frac{1}{N_{k+1,1}}}_{\substack{\text{additional} \\ \text{factor}}}\\
 &= \frac{(-1)^{N_{k+1,0}}}{\prod\limits_{i=1}^{k+1} i^{N_{k+1,i}}\, N_{k+1,i}!}\\
 &= Q_{N_{k+1,0},\ldots,N_{k+1,k+1}} \quad.
\end{align*}

We proceed likewise with the other $Q_{i_j}$'s, which will be factored in the terms coming from $K_2$ when $x_u=i_j$ and also when $k=i_j$. The $i_j$'s are in fact the only possible values for $x_u$ since, at a fixed partition of $k$, the values of $x_u$ giving non-zero terms in the sum are those corresponding to non-zero $N_{k,x_u}$. This time, it is the double summation over $u$ and $v$ in $K_2$ which will allow us to transform the sum over $\sigma'\in\mathfrak{S}_k$ into a sum over $\mathfrak{S}_{k+1}$ without additional normalization factor. To each permutation $\sigma'\in\mathfrak{S}_k$, we can associate
 the following permutations $\sigma'_{k+1,u,v}\in\mathfrak{S}_{k+1}$:
\begin{equation}
 \left\{ \begin{array}{l l}
  \sigma'_{k+1,u,v}(\sigma'^{-1}(v))=k+1\ \text{if } v\neq k+1\\
  \sigma'_{k+1,u,v}(L_{i_j})=u\\
  \sigma'_{k+1,u,v}(k+1)=v\\
  \sigma'_{k+1,u,v}(l)=\sigma'(l) \ \text{for all } l\in\{1,\ldots,k\} \setminus \{\sigma'^{-1}(v),L_{i_j}\}
 \end{array} \right. \quad.
\end{equation}
These permutations are pairwise distinct when we consider all the possible values for $u$ and $v$ and all the permutations $\sigma'\in\mathfrak{S}_k$.  In $K_2$, the double summation on $(u,v)$ contains $\frac{k(k+1)}{2}$  values and the sum over  $\sigma'\in\mathfrak{S}_k$ contains $(k-1)!$ different permutations as  $\sigma'(L_{i_j})=u$ is fixed. Therefore, we arrive at $\frac{(k-1)!\,k(k+1)}{2}=\frac{(k+1)!}{2}$ distinct permutations in $\mathfrak{S}_{k+1}$. This number has to be multiplied by two, since, as already noticed, the $(i_j+1)$-cycle trace containing $C_{k+1}^\dagger$ actually counts as two (the previous $\frac{(k+1)!}{2}$  permutations can be composed with transposition $\tau_{u,v}$ to account for the symmetrized $C$-matrix product). So, we have a total of $(k+1)!$ distinct permutations, which means that we have all the permutations belonging to $\mathfrak{S}_{k+1}$.
Regarding the extension of the summation over $\sigma$ to $\mathfrak{S}_{k+1}$, we have to divide by the number of ways to interleave the new $(i_j+1)$-cycle trace (containing $C_{k+1}^\dagger$) with the  $(N_{k+1,i_j+1}-1)$ other traces, but also by a factor $(i_j+1)$, which arises from the  invariance of the $(i_j+1)$-cycle trace containing $C_{k+1}^\dagger$  under circular permutation. We obtain as required:
\begin{align*}
 &\!\!\!\!\!\!\!\!\!\!\!\!\!\!\!\!\!\!\!\!Q_{i_j} =  \underbrace{-Q_{N_{k+1,0}-1,N_{k+1,1},\ldots,N_{k+1,i_j-1},N_{k+1,i_j}+1,N_{k+1,i_j+1}-1,N_{k+1,i_j+2},\ldots,N_{k+1,k}}(N_{k+1,i_j}+1)\,i_j}_{\text{factor $-Q_{N_{k,0}^{i_j},\ldots,N_{k,k}^{i_j}}N_{k,i_j}^{i_j}\, i_j$ appearing in $K_2$}}\times\underbrace{\frac{1}{N_{k+1,i_j+1}(i_j+1)}}_{\substack{\text{additional} \\ \text{factor}}} \\
 &\!\!\!\!\!\!\!\!\!\!= \frac{(-1)^{N_{k+1,0}}}{\prod\limits_{i=1}^{k+1} i^{N_{k+1,i}}\, N_{k+1,i}!}\\
 &\!\!\!\!\!\!\!\!\!\!=Q_{N_{k+1,0},\ldots,N_{k+1,k+1}} \quad.
\end{align*}

To resume our achievements: the sum over partitions of $k$ has been transformed into a sum over partitions of $k+1$, and all the other sums have been condensed into a double sum over $\mathfrak{S}_{k+1}$ while the normalization factors turned into the expected $Q_{N_{k+1,0},\ldots,N_{k+1,k+1}}$ factors. All the terms in $K_1$ and $K_2$ have been consumed along the way.
This  implies that $\langle \Phi_1\wedge\cdots\wedge\Phi_{k+1} |  \Phi'_1\wedge\cdots\wedge\Phi'_{k+1} \rangle=K_1+K_2$ verifies our induction property for Step $(k+1)$.  This concludes our proof by induction of the general APG overlap formula.\\

\end{document}